\documentclass[a4paper,fleqn,usenatbib]{mnras}

\usepackage{newtxtext,newtxmath}
\usepackage{gensymb}

\usepackage[T1]{fontenc}
\usepackage{ae,aecompl}

\usepackage{anyfontsize}

\usepackage{siunitx}

\usepackage{graphicx}	
\usepackage{amsmath}	
\usepackage{nccmath}
\usepackage{longtable}
\usepackage{arydshln}
\usepackage{booktabs}
\usepackage{tabularx}
\usepackage{supertabular}
\usepackage{enumitem}
\setlist{nosep,leftmargin=*}

\title[Multi-scale environmental quenching]{Large-scale and local environmental drivers of quenching: \\ tracing H$\alpha$ concentration in X-ray and optical galaxy groups}

\author[S.\ Barsanti et al.]{Stefania Barsanti,$^{1,2,3}$\thanks{E-mail: stefania.barsanti@sydney.edu.au}
Di Wang,$^{1,2}$
Matthew Colless,$^{3,2}$
Ang Liu,$^{4,5}$
Esra Bulbul,$^{5}$
\newauthor
Matt S. Owers,$^{6,2,7}$
Scott M. Croom,$^{1,2}$
Benedetta Vulcani,$^{8}$
Julia J. Bryant,$^{1,2,9}$
Yifan Mai,$^{10,2,7}$
\newauthor
Sree Oh,$^{11}$
Andrei Ristea,$^{12,13}$
Sarah M. Sweet,$^{14,2}$ 
Jesse van de Sande$^{15,2}$
\vspace{0.4cm}
\\
$^{1}$Sydney Institute for Astronomy (SIfA), School of Physics, The University of Sydney, NSW 2006, Australia\\
$^{2}$ARC Centre of Excellence for All Sky Astrophysics in 3 Dimensions (ASTRO 3D), Australia\\
$^{3}$Research School of Astronomy and Astrophysics, Australian National University, Canberra, ACT 2611, Australia\\
$^{4}$Institute for Frontiers in Astronomy and Astrophysics, Beijing Normal University, Beijing 102206, China\\
$^{5}$Max Planck Institute for Extraterrestrial Physics, Giessenbachstrasse 1, 85748, Garching, Germany\\
$^{6}$School of Mathematical and Physical Sciences, Macquarie University, Sydney, NSW 2109, Australia\\
$^{7}$Astrophysics and Space Technologies Research Centre, Macquarie University, Sydney, NSW 2109, Australia\\
$^{8}$INAF – Osservatorio Astronomico di Padova, Vicolo Osservatorio 5, 35122, Padova, Italy\\
$^{9}$Australis-USydney, School of Physics, University of Sydney, NSW 2006, Australia\\
$^{10}$Australian Astronomical Optics, Macquarie University, Sydney, NSW 2109, Australia\\
$^{11}$Department of Astronomy and Yonsei University Observatory, Yonsei University, Seoul, 03722, Republic of Korea\\
$^{12}$Centre for Astrophysics and Supercomputing, Swinburne University of Technology, Hawthorn, VIC 3122, Australia\\
$^{13}$ARC Centre of Excellence in Optical Microcombs for Breakthrough Science (COMBS), Australia\\
$^{14}$School of Mathematics and Physics, University of Queensland, Brisbane, QLD 4072, Australia\\
$^{15}$School of Physics, University of New South Wales, Sydney, NSW 2052, Australia\\
}

\date{Accepted 2026 May 05. Received 2026 April 01; in original form 2025 October 16.}

\pubyear{2026}

\begin{document}
\label{firstpage}
\pagerange{\pageref{firstpage}--\pageref{lastpage}}
\maketitle

\begin{abstract}

To explore the environmental mechanisms causing quenching in nearby star-forming galaxies, we study the variation with local and large-scale environments of a star formation concentration index, C-index $\equiv\log{(r_{50,{\rm H}\alpha}/r_{50,\rm cont}})$, that traces the spatially-resolved distribution of H$\alpha$ emission. Our analysis combines (i)~GAMA spectroscopic redshift survey data to optically select galaxy groups and reconstruct the cosmic web, (ii)~eROSITA data to identify X-ray-emitting groups, and (iii)~SAMI Galaxy Survey data to characterise spatially-resolved star formation. We find that galaxies in X-ray+optical groups exhibit the lowest median C-index and the highest fraction of centrally-concentrated star-forming galaxies relative to optical groups and the field (independently of group or stellar mass). Star-forming galaxies in more X-ray luminous groups at fixed dynamical mass show more concentrated star formation. At large scales, nodes show the lowest median C-index and the highest fraction of centrally-concentrated star-forming galaxies relative to filaments and voids, which have similar C-index distributions. C-index correlates most strongly with the distance to the closest node, leaving no significant role for other local or large-scale environment metrics. Finally, regular star-forming galaxies tend to have spins aligned parallel to filaments, consistent with smooth gas accretion, while centrally-concentrated galaxies tend have spins aligned perpendicular to filaments, likely driven by mergers and associated with bulge growth. These results suggest that multi-scale environmental processes, i.e. locally and at large-scale, act to concentrate star formation toward galaxy centres, via gas-related mechanisms in nodes and ram-pressure stripping in X-ray+optical groups.
\end{abstract}



\section{Introduction}
\label{Introduction}

The quenching of star formation in galaxies is a critical process in understanding galaxy evolution, transforming actively star-forming galaxies into quiescent populations. In some galaxies, star formation is suppressed first in the central regions of a galaxy and then extends progressively to its outer regions. This inside-out quenching mode is commonly associated with internal processes that preferentially affect the galaxy core, such as AGN feedback, stellar feedback or gas exhaustion \citep{Tacchella2015,Belfiore2017}. In other galaxies, the suppression of star formation begins in the galaxy’s outskirts and progressively moves inward toward the central regions \citep{Ellison2018}. This outside-in quenching mode is typically driven by external environmental mechanisms.

Advances in integral field spectroscopy surveys have provided crucial insights into spatially-resolved quenching. One key metric is the star formation concentration index (C-index), a parameter that quantifies the spatially-resolved distribution of H$\alpha$ emission that is defined by
\begin{equation}\label{eqn:C-index}
     \log(C_{{\rm H}\alpha})=\log\left(\frac{r_{50,{\rm H}\alpha}}{r_{50,\rm cont}}\right)
\end{equation}
where $r_{50,{\rm H}\alpha}$ is the radius enclosing 50\% of the H$\alpha$ emission (tracing recent star formation) and $r_{50,\rm cont}$ is the radius enclosing 50\% of the stellar continuum light (tracing the overall stellar distribution). A lower C-index indicates more centrally concentrated star formation, which can be a signature of outside-in quenching. Studies leveraging the SAMI Galaxy Survey have demonstrated the correlation between concentrated star formation and environment metrics. \citet{Schaefer2016} showed that the fraction of galaxies with centrally concentrated star formation increases with local galaxy density, pointing to quenching occurring via the outside-in mode in dense environments. \citet{Schaefer2019} studied spatially-resolved star formation in galaxy groups, finding signatures of environmental quenching in high-mass groups. \citet{Owers2019} found signs of outside-in quenching by ram-pressure stripping within the eight SAMI galaxy clusters. \citet{Wang2022} found that the fraction of concentrated star-forming galaxies increases as halo mass increases, suggesting outside-in quenching occurring in both groups and clusters. In agreement, \citet{Pan2025} concluded that C-index is sensitive to environmental effects using the MaNGA Galaxy Survey.

Ram-pressure stripping stands out as a dominant outside-in quenching process, where a galaxy's cold gas reservoir is stripped by the pressure of the hot intergalactic medium (called intragroup medium - IGM - within galaxy groups) as the galaxy moves through it, preventing star formation in the outer disc \citep{Gunn1972}. Other processes contributing to outside-in quenching include tidal interactions, where gravitational forces disrupt the outer regions of a galaxy, reducing gas densities and star formation activity \citep{Merritt1984}, and strangulation, where the removal or depletion of gas reservoirs in the outer regions over time limits the supply of fuel for star formation \citep{Larson1980}. However, the relative importance of these mechanisms varies across different environments. Ram-pressure stripping plays a major role in clusters \citep{Boselli2006,Chung2017,Vulcani2022}, and it has also been found to be effective in groups \citep{Bureau2002,McConnachie2007,Brown2017}. Ram-pressure stripping can take place in various environments as long as a galaxy moves at a sufficiently high velocity through a medium dense enough (regardless of its temperature) to overcome the gravitational potential binding the galaxy's interstellar medium, efficiently stripping the gas and quenching star formation activity \citep{Boselli2022}. 

X-ray observations of halos are able to provide a direct observational window into ram-pressure stripping effects, as the X-ray emission traces the hot and diffuse IGM. Galaxies in these X-ray-detected groups or clusters often exhibit signatures of ram-pressure stripping, such as trailing gas tails \citep{Chung2007,Poggianti2017}, asymmetric gas distributions \citep{Kenney2004,Rasmussen2006}, and radial gradients in star formation activity \citep{Boselli2006,Crowl2008}. In particular, \citet{Rasmussen2006} finds that the degree of gas depletion and star formation quenching correlates with the X-ray luminosity of the group, suggesting a stronger impact of ram-pressure stripping in groups with denser IGM. Thus, X-ray-detected halos can play a crucial role in understanding the impact of ram-pressure stripping on galaxy evolution.

X-ray-detected halos are found at the nodes of the cosmic web \citep{Eckert2015,Popesso2024}. However, as galaxies migrate along filaments toward dense cluster cores, they encounter environmental mechanisms that influence their gas reservoirs and star formation activity. Ram-pressure stripping has also been found to act within filaments and nodes \citep{BenitezLlambay2013,Vulcani2018,Vulcani2021}, highlighting the role of the cosmic web as a potential precursor to cluster-induced quenching \citep{Kraljic2018}. 

Moreover, cosmic web location has an effect on galaxy kinematics, influencing the orientation and the amplitude of galaxy spin \citep{Barsanti2025}. From large-scale cosmological hydrodynamical simulations, a parallel alignment between the galaxy's spin axis and the orientation of the closest filament is associated with galaxies that have formed through smooth gas accretion along filaments, which preserves angular momentum aligned with the filament. In contrast, perpendicular galaxy spin--filament alignments suggest that these galaxies have undergone mergers, disrupting the coherent gas accretion and altering their spin orientation \citep{Dubois2014,Welker2014}. Such simulations are in agreement with observations, where perpendicular galaxy spin--filament alignments are more frequent for more massive galaxies \citep{Welker2020} and for galaxies with higher bulge masses \citep{Barsanti2022}. The cosmic web thus plays an active role in shaping galaxy evolution \citep{Nandi2025}.

This work explores star formation concentration for nearby star-forming galaxies in local and large-scale environments to understand the physical processes that regulate spatially-resolved quenching across different environmental scales. We take advantage of (i)~the GAMA spectroscopic redshift survey for the optical selection of galaxy groups and for reconstructing the cosmic web; (ii)~the eROSITA data within the GAMA regions to identify optically-selected groups with X-ray detections; and (iii)~the SAMI Galaxy Survey for spatially-resolved galaxy properties. 

This paper is structured as follows: Section~\ref{Dataset} describes our dataset; Section~\ref{Method} details the characterisation of local environments, the cosmic web reconstruction, and galaxy properties; Section~\ref{Results} presents our results on the C-index of star-forming galaxies as a function of local and large-scale environments; Section~\ref{Discussion} compares our findings to previous studies and discusses physical interpretations; our conclusions are stated in Section~\ref{Summary and conclusions}. Throughout this work, we assume $\Omega_{m}=0.3$, $\Omega_{\Lambda}=0.7$, and $H_0=70$\,km\,s$^{-1}$\,Mpc$^{-1}$.

\section{Dataset}
\label{Dataset}

\subsection{The GAMA survey}
\label{GAMA galaxy survey}

The Galaxy And Mass Assembly survey (GAMA; \citealp{Driver2011,Hopkins2013,Baldry2018,Driver2022}) is a spectroscopic and photometric survey of $\sim$300,000 galaxies with $r \le 19.8$\,mag that covers $\sim$286\,deg$^{2}$ in 5 regions called G02, G09, G12, G15 and G23. The redshift range of the GAMA sample is $0<z<0.5$, with a median redshift of $z\sim0.25$. Most of the spectroscopic data were obtained using the AAOmega multi-object spectrograph at the Anglo-Australian Telescope, although GAMA also incorporates previous spectroscopic surveys such as SDSS \citep{York2000}, 2dFGRS \citep{Colless2001,Colless2003}, the Millennium Galaxy Catalogue \citep{Driver2005} and WiggleZ \citep{Drinkwater2010}. 

The GAMA catalogue of galaxy groups is built using a friends-of-friends algorithm that examines both radial and projected comoving distances to assess overlapping galactic halo memberships \citep{Robotham2011}.  The group catalog contains 23654
groups (each with $\geq$2 members) and 184081 galaxies from the G09, G12 and G15 regions observed down to $r < 19.8$ mag.

The GAMA survey’s extensive and highly complete spectroscopic redshift data (98.5\% in the equatorial regions; \citealp{Liske2015}), paired with its wide area coverage, high spatial resolution, and broad wavelength range, also make it an exceptional dataset for mapping the cosmic web.

\subsection{eROSITA/eFEDS and eROSITA/eRASS1 data}
\label{eROSITA eFEDS and eRASS1 data}

The eROSITA Final Equatorial-Depth Survey (eFEDS) is a large public X-ray survey conducted during the performance verification phase of the eROSITA instrument on the SRG (Spectrum-Roentgen-Gamma) satellite \citep{Predehl2021,Sunyaev2021}. eFEDS covers an equatorial area of approximately 140 square degrees, including the GAMA G09 region. The X-ray sources are classified as point-like or extended based on their X-ray morphology \citep{Brunner2022}; they are further categorised, using multi-wavelength data, into Galactic sources, AGN, individual galaxies, galaxy groups and clusters \citep{Bulbul2022,Liu2022}. We employ the eROSITA/eFEDS group and cluster catalogue from \citet{Liu2022}, which includes over 500 extended objects up to $z\sim1$. The optical confirmation and redshift determination of galaxy clusters and groups in the eFEDS field is detailed in \citet{Klein2022}. X-ray luminosities are reported within 300\,kpc and 500\,kpc.  

The first all-sky X-ray survey conducted by eROSITA is eRASS1 \citep{Merloni2024}. It represents the first complete scan of the eight planned X-ray surveys and was finished in the first six months of the mission's full operation. We employ the eROSITA/eRASS1 group and cluster catalogue public V3.2 release from \citet{Bulbul2024}, which includes 12,247 optically-confirmed halos detected in the 0.2--2.3\,keV range as extended X-ray sources over a region of 13,116 square degrees in the western Galactic hemisphere of the sky, covering also the GAMA G12 and G15 regions. The redshift range of the halos is $0.003 \leq z \leq 1.32$. The optical identification and properties of the eRASS1 galaxy clusters are discussed in \citet{Kluge2024}.

Because of the relatively large point-spread function of eROSITA, high-redshift galaxy clusters or compact nearby groups hosting bright active galactic nuclei can sometimes be misclassified as point sources by the source detection algorithms. Thus, we also examine catalogues of groups that may have been disguised as point sources, as presented by \citet{Bulbul2022} for eFEDS and by \citet{Balzer2025} for eRASS1. These catalogues include 346 eFEDS groups in the range 
$0.1<z<1.3$ and 8,347 groups from eRASS1 spanning $0.05<z<1.1$. In cases where objects appear in both the eFEDS and eRASS1 catalogues, we adopt the eFEDS entries, as the eFEDS observations are deeper for source characterisation.

\subsection{The SAMI Galaxy Survey}
\label{SAMI galaxy survey}

The SAMI Galaxy Survey is a spatially-resolved spectroscopic survey of 3068 galaxies with stellar masses $\log(M_{\star}/M_{\odot})$\,=\,8--12 and redshifts $0.004<z\leq0.115$ \citep{Bryant2015,Croom2021}. Most of the SAMI targets belong to the three GAMA equatorial fields (G09, G12 and G15). Eight massive clusters with virial masses $14.5\leq\log(M_{200}/M)\leq15.2$ were also observed \citep{Owers2017}, but are excluded from our analysis since we focus on galaxy groups. 

The Sydney--AAO Multi-object Integral-field spectrograph (SAMI) was mounted on the 3.9\,m Anglo-Australian Telescope \citep{Croom2012}. The instrument had 13 fused optical fibre bundles (hexabundles), each containing 61 fibres of 1.6\,arcsec diameter so that each integral field unit had a 15\,arcsec diameter \citep{Bland2011,Bryant2014}. The SAMI fibres fed the two arms of the AAOmega spectrograph \citep{Sharp2006}. The SAMI Galaxy Survey used the 580V grating in the blue arm, giving a resolving power of $R$=1812 and wavelength coverage of 3700--5700\,\AA, and the 1000R grating in the red arm, giving a resolving power of $R$=4263 over the range 6300--7400\,\AA. The median full-width-at-half-maximum values for each arm are FWHM$_{\rm blue}=2.65$\,\AA\ and FWHM$_{\rm red}=1.61$\,\AA\ \citep{vandeSande2017}. The SAMI data-cubes comprise a grid of 0.5$\times$0.5\,arcsec spaxels, with blue and red spectra having pixels corresponding to 1.05\,\AA\ and 0.60\,\AA\ respectively.

\begin{figure}
\centering
\includegraphics[width=\columnwidth]{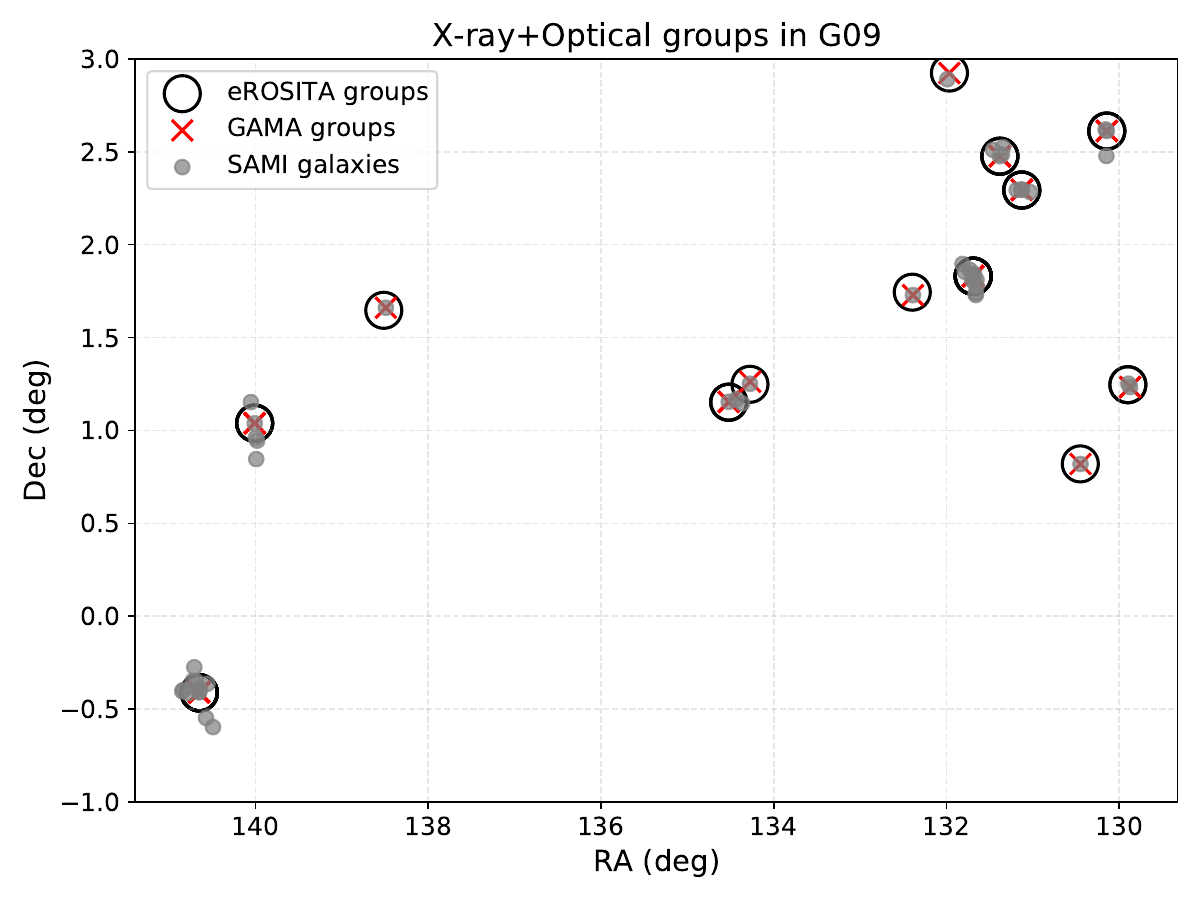}
\caption{X-ray+Optical groups in G09 with SAMI galaxies (gray dots) overlapping the optical group centre (red cross) and the X-ray group centre (black circle).}
\label{SAMI_GAMA_eROSITA_G09}
\end{figure}

\section{Method}
\label{Method}

\subsection{Characterisation of local environments}
\label{Characterisation of local environments}

To understand the impact of outside-in star formation quenching in various galaxy habitats, we characterise local environments by three categories: (i)~X-ray+Optical groups: optically-selected galaxy groups with X-ray detections; (ii)~Optical groups: optically-selected galaxy groups without X-ray detections; and (iii)~Field: optically-selected galaxies not in any group. We capitalise these categories to indicate these specific meanings.

An optically-selected group is defined as a GAMA group from the catalogue of \citet{Robotham2011} in the equatorial G09, G12 or G15 fields (GAMA, eROSITA and SAMI overlapping regions), with group edge~>~0.70 (i.e.\ the group centre is recovered and at least 70\% of the group’s area is inside the survey), at least 5 member galaxies, redshift range $z<0.13$ (SAMI upper limit), and $12<\log(M_{200}/M_{\odot})<15$. The group mass $M_{200}$ is defined as the mass of a spherical halo with a mean density that is 200 times the critical cosmic density at the halo redshift. The group edge threshold adopted provides an optimal balance between maximising the number of available groups and ensuring that their membership and environmental measures are not strongly biased by survey edge effects. There are 1315 optically-selected groups that match these criteria. 

We adopt a $120^{\prime\prime}$ (2 arcmin) matching radius when cross-matching optical groups with eROSITA-based catalogues. This choice reflects the relatively large eROSITA PSF (HEW $\sim26^{\prime\prime}$ on-axis, increasing off-axis) and the fact that X-ray and optical centres of groups can be offset by tens of arcseconds due to substructure or AGN contamination. Similar matching scales (1–2 arcmin) have been adopted in other eROSITA cluster studies \citep{Liu2021, Bulbul2022, Seppi2022}. There are 79 X-ray+Optical groups versus 1236 Optical groups with no X-ray detection. Out of these, there are 20 X-ray+Optical groups containing 77 galaxies with SAMI spatially-resolved data and 187 Optical groups containing 467 SAMI galaxies. Figure~\ref{SAMI_GAMA_eROSITA_G09} shows the X-ray+Optical groups in G09 (populated by 13 out of 20 of these halos) with SAMI galaxies (gray dots) overlapping the optical group centre (red cross) and the X-ray group centre (black circle). We assess the sensitivity of the optical and X-ray group matching to the adopted angular radius by repeating the analysis with matching radii of $60^{\prime\prime}$ and $180^{\prime\prime}$ in Appendix~\ref{Sensitivity of the optical and Xray group matching}. The conclusions of this work remain unchanged.

\begin{figure*}
    \centering
    {\includegraphics[width=0.49\textwidth]{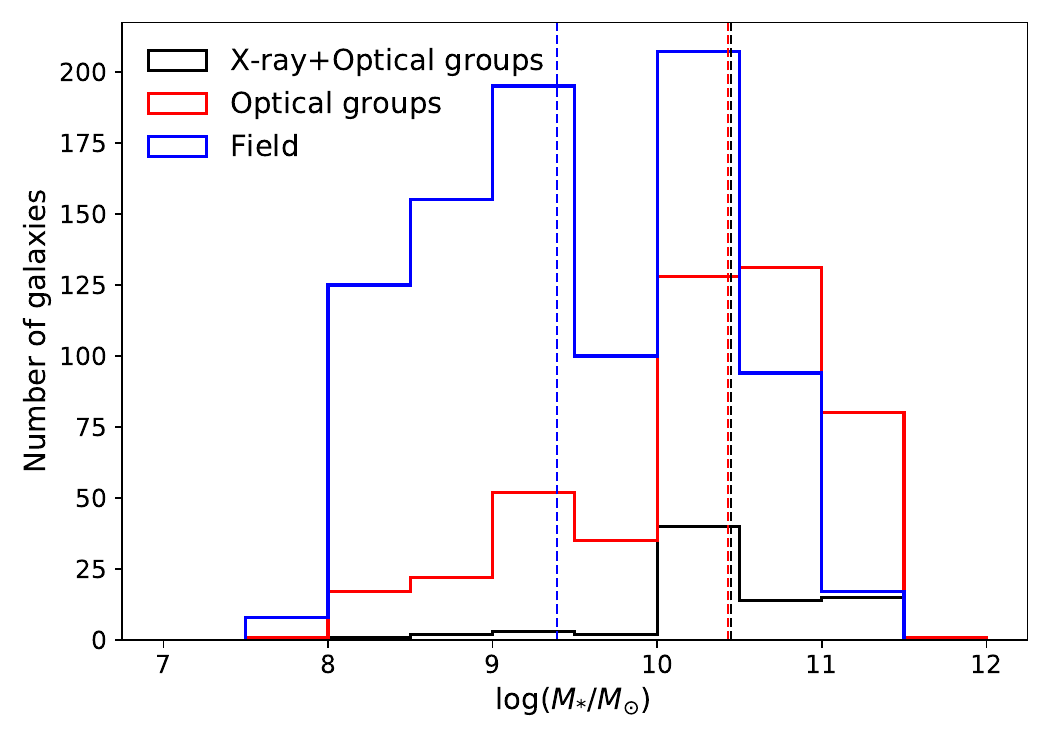}} 
    {\includegraphics[width=0.49\textwidth]{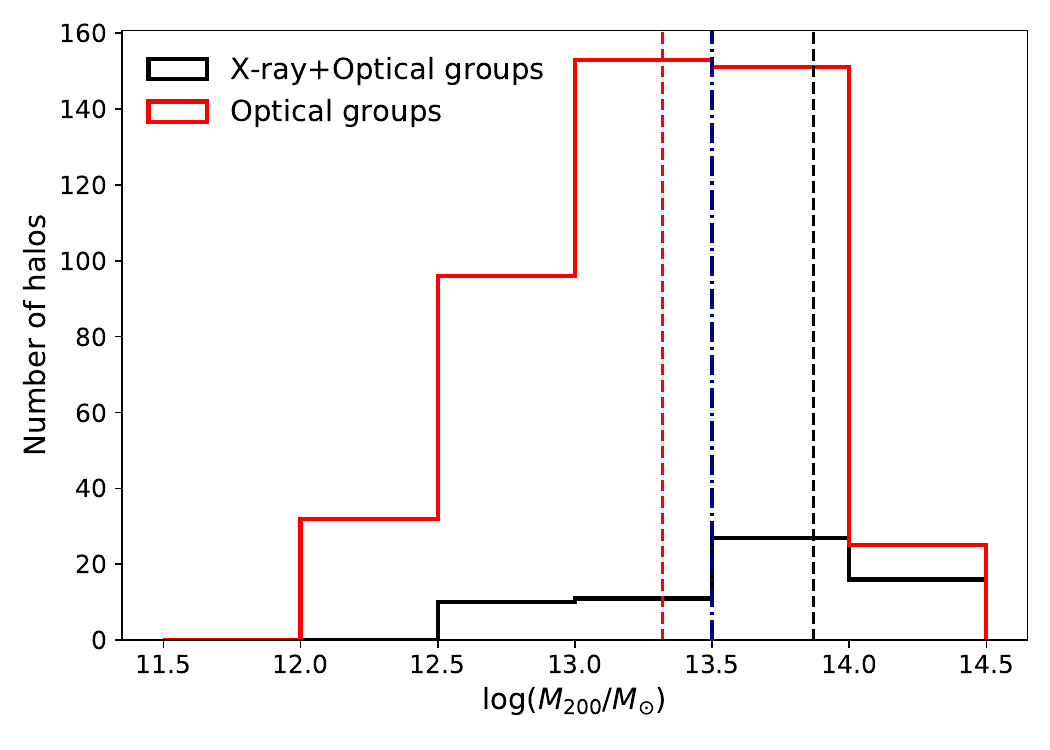}}
    \caption{Left: stellar mass distributions for the local environments. Right: halo mass distributions for X-ray+Optical groups (black) and Optical groups (red). Dashed lines in both panels are medians; the dash-dot line (right panel) is the $M_{200}$ above which the distributions are statistically consistent.}
    \label{M200_Mstar_histograms}
\end{figure*} 

\begin{figure*}
\includegraphics[width=\textwidth]{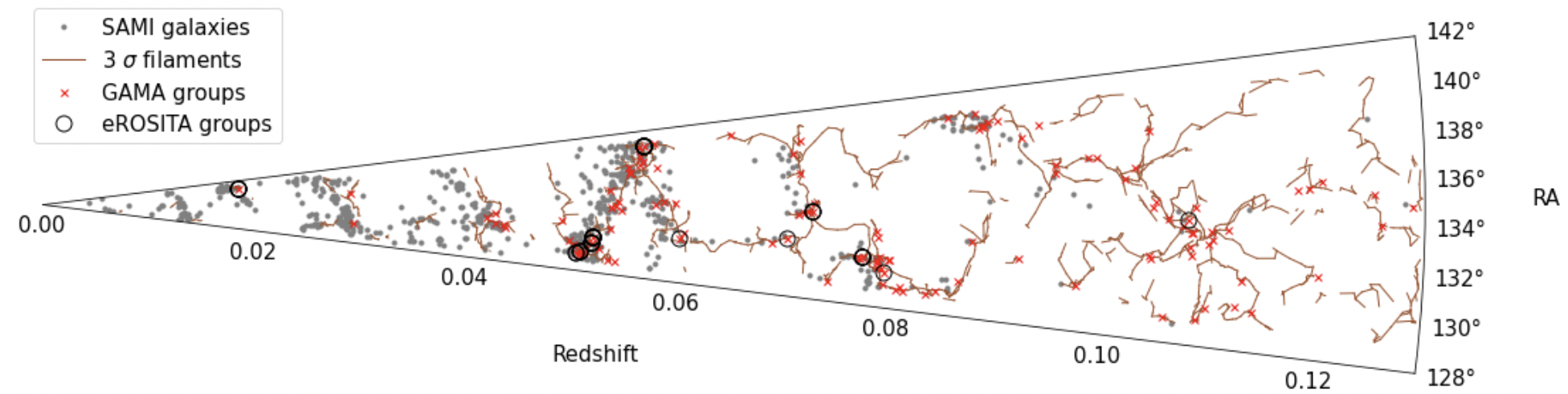}
\caption{Projected network of cosmic filaments (brown lines) for the GAMA G09 region. Red crosses are the GAMA optical group centres, black circles are the cross-matched eROSITA X-ray centres and gray points are SAMI galaxies.}
\label{PolarPlotSAMIGAMAFilaments}
\end{figure*}

The Field comprises 903 SAMI galaxies. Stellar mass ($M_\star$) distributions for SAMI galaxies in the three local environments are reported in the left panel of Figure~\ref{M200_Mstar_histograms}. Galaxy stellar masses are measured from K-corrected $g$$-$$i$ colours and $i$-band magnitudes for the SAMI Galaxy Survey \citep{Bryant2015}. According to a two-sample Kolmogorov-Smirnov (K-S) test \citep{Lederman1984}, the Field is populated by significantly lower-$M_{\star}$ galaxies with respect to Optical groups ($p_{\rm KS}=10^{-20}$) and X-ray+Optical groups ($p_{\rm KS}=10^{-39}$), while the distributions for X-ray+Optical groups and Optical groups are marginally different ($p_{\rm KS}=0.027$). 

The spatially-resolved H$\alpha$ distribution is also explored as a function of local environment metrics, namely local galaxy density and halo mass. Local galaxy density is measured as the fifth-nearest neighbour surface density $\Sigma_5$ \citep{Brough2013}. We use the GAMA galaxy group catalogue built with a friend-of-friends algorithm \citep{Robotham2011} to estimate halo masses from group velocity dispersions and projected distances. The right panel of Figure~\ref{M200_Mstar_histograms} shows the halo mass distributions for X-ray+Optical groups and Optical groups. X-ray+Optical groups tend to be more massive than Optical groups. According to the two-sample K-S test, the halo mass distributions are not significantly different for masses above $\log(M_{200}/M_{\odot})=13.5$ (dash-dot line), allowing us to study galaxy properties in these local habitats for a consistent halo mass range beyond this cut.

Since halo masses derived from group velocity dispersions become increasingly uncertain for systems with small membership, we cross-check $M_{200}$ estimates for the X-ray+Optical groups against $M_{500}$ values inferred from the X-ray luminosity, using the scaling relations of \citet{Chiu2022}. We find good agreement with $M_{200}\simeq1.34\,M_{500}$, as expected for group-scale haloes at low redshift \citep{Hu2003}. This provides reassurance that $M_{200}$ estimates for our groups sample are robust.

\subsection{Characterisation of large-scale environments}
\label{Large-scale environments}

We use the cosmic web reconstruction for the G09, G12 and G15 regions from \citet{Barsanti2022}, based on the public Discrete Persistent Structure Extractor code ({\sc DisPerSE}; \citealp{Sousbie2011a,Sousbie2011b}). {\sc DisPerSE} is a parameter-free, scale-independent algorithm grounded in discrete Morse theory and persistence theory. It identifies voids, walls, and filaments as distinct components of the cosmic web using galaxies' sky positions and spectroscopic redshifts. These structures correspond to a topological segmentation of space defined by the gradient of the Morse function: voids are 3D regions around local minima, walls are 2D surfaces separating adjacent voids, and filaments are 1D connections linking maxima. The gradients of the Morse function in {\sc DisPerSE} mathematically describe flow lines connecting minima, saddles, and maxima of the density field. These correspond directly to the dynamical gradients of the tidal field, which govern anisotropic collapse. Maxima/minima/saddles in Morse theory map onto the tidal eigenvalue picture of nodes, voids, sheets, and filaments. So {\sc DisPerSE} essentially extracts the observable/topological cosmic web skeleton that has its physical origin in the tidal field.

The identification of cosmic filaments can be impacted by the ‘Fingers of God’ effect (FoG; \citealp{Jackson1972}), where galaxy groups and clusters appear elongated along the line of sight in redshift space due to internal velocity dispersions. Such distortions may lead to the detection of spurious filaments. However, \citet{Barsanti2022} demonstrate that correcting for FoG has only a minor influence on the resulting cosmic web structure for the SAMI Galaxy Survey due to the SAMI volume probing only the nearby region of the GAMA survey at $0< z < 0.12$, where the number of rich and massive groups, for which the FoG effect is most important, is very limited \citep{Barsanti2018}.

\begin{figure*}
    \centering
    {\includegraphics[width=0.49\textwidth]{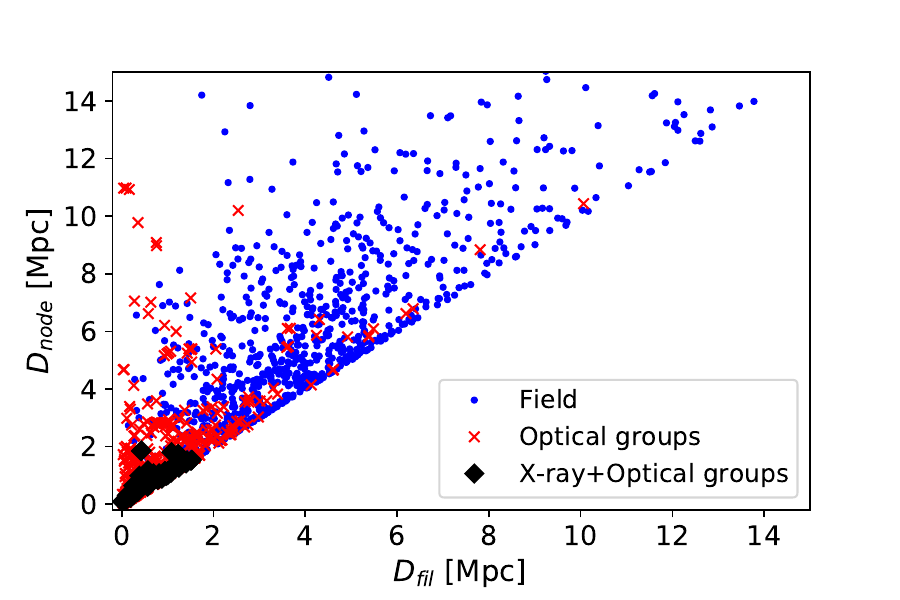}} 
    {\includegraphics[width=0.49\textwidth]{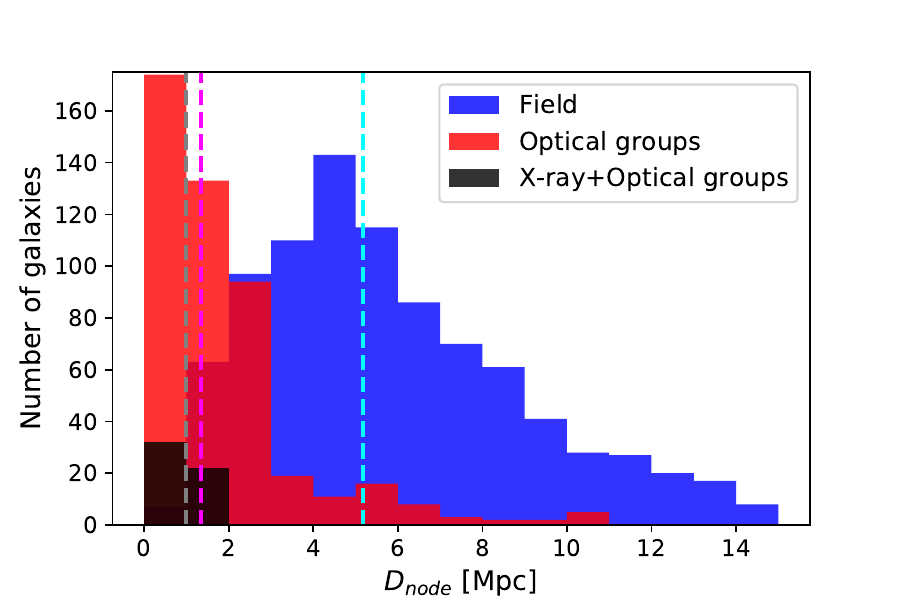}}
    \caption{Left panel: galaxies of local environments within the $D_{\rm fil}$ versus $D_{\rm node}$ plane. Right panel: $D_{\rm node}$ distributions for galaxies in X-ray+Optical groups (black), Optical groups (red) and Field (blue). Dashed lines are medians. X-ray+Optical groups are found closest to nodes. }
    \label{LocalEnvironments_CW}
\end{figure*} 

For the galaxy distribution, \citet{Barsanti2022} use 35,882 GAMA galaxies from DR3 \citep{Baldry2018} with secure redshifts and stellar masses in the SAMI redshift range ($0<z<0.13$) within the G09, G12 and G15 regions. Most galaxies have stellar masses between $10^{8}\,M_{\odot}$ and $10^{12}\,M_{\odot}$; the median stellar mass and redshift are $M_\star \sim 10^{9.7}\,M_{\odot}$ and $z \sim 0.09$. {\sc DisPerSE} is then run with a 3$\sigma$ persistence threshold to identify the most significant structures. The 3D filamentary structure is shown as an interactive plot at \href{https://skfb.ly/o9MXz}{this URL}; Figure~\ref{PolarPlotSAMIGAMAFilaments} shows the projected network of filaments (brown lines) for the G09 region. The centres of the GAMA optical groups (red crosses) are found at the intersections between filaments; the cross-matched eROSITA X-ray centres are marked by black circles and gray points show SAMI galaxies. We measure the distance from each SAMI galaxy to the closest cosmic filament ($D_{\rm fil}$) and to the closest node ($D_{\rm node}$) using the smallest 3D Euclidean distance (a schematic representation of these cosmic web metrics is shown in Figure~2 of \citealp{Barsanti2025}); by construction, $D_{\rm fil}$ is always smaller than $D_{\rm node}$. Following \citet{Barsanti2025}, we also classify galaxies by the cosmic structure they reside within: we identify galaxies with $D_{\rm node}<1$\,Mpc as belonging to Nodes, galaxies with $D_{\rm node}>1$\,Mpc and $D_{\rm fil}<1.5$\,Mpc as belonging to Filaments, and the remaining galaxies as belonging to Voids (capitalised to indicate these definitions). 

We clarify that the galaxy catalogue used as input to {\sc DisPerSE} is restricted to the GAMA survey footprint (G09, G12, G15), with regions outside the angular mask explicitly excluded from the density field. This prevents {\sc DisPerSE} from interpreting sharp survey boundaries as physical density gradients or filamentary structures. The {\sc DisPerSE} reconstruction further adopts periodic boundary conditions for the 3D Delaunay tessellation and smooth boundary conditions for the density field analysis, which regularise the density field near survey boundaries. As an additional robustness check, we test for residual edge effects by excluding galaxies within 3$\,$Mpc (i.e. $\geq 2\times$ the 1.5 Mpc filament-association scale) of the survey boundaries, and find no impact on our conclusions. We discuss the impact of flux-limited selection and redshift-dependent sampling on the cosmic web analysis in Appendix~\ref{Impact of flux-limited selection and redshift-dependent sampling}, showing that inherent GAMA and SAMI selection biases do not affect our results. Finally, we discuss the sensitivity of the cosmic web reconstruction to the chosen {\sc DisPerSE} persistence threshold in Appendix~\ref{Sensitivity of the Cosmic Web Reconstruction to the Persistence Threshold}.

In order to better understand the locations of galaxies belonging to different local environments (X-ray+Optical groups, Optical groups, and Field; see Section~\ref{Characterisation of local environments}) within the reconstructed cosmic web, we show in the left panel of Figure~\ref{LocalEnvironments_CW} the distances of individual galaxies to the nearest node as a function of their distance to the nearest filament. The cosmic web classification is independent from the characterisation of local environments, so examining potential overlaps between these environments, as well as analysing each environemntal set separately, provides important and complementary insights into how local and large-scale environments shape star formation. There is a lack of Field galaxies at low values of $D_{\rm fil}-D_{\rm node}$, while galaxies in Optical groups tend to be concentrated closer to nodes and filaments, and galaxies in X-ray+Optical groups are found only at $D_{\rm fil}<2$\,Mpc and $D_{\rm node}<2$\,Mpc. The right panel of Figure~\ref{LocalEnvironments_CW} displays the $D_{\rm node}$ distributions for X-ray+Optical groups (black), Optical groups (red) and Field (blue). The median distance to the closest node (dashed line) increases from X+ray+Optical groups to Optical groups to the Field.

\subsection{Characterisation of star-forming galaxies}
\label{Characterisation of star-forming galaxies}

\begin{figure}
    \centering
    {\includegraphics[width=\columnwidth]{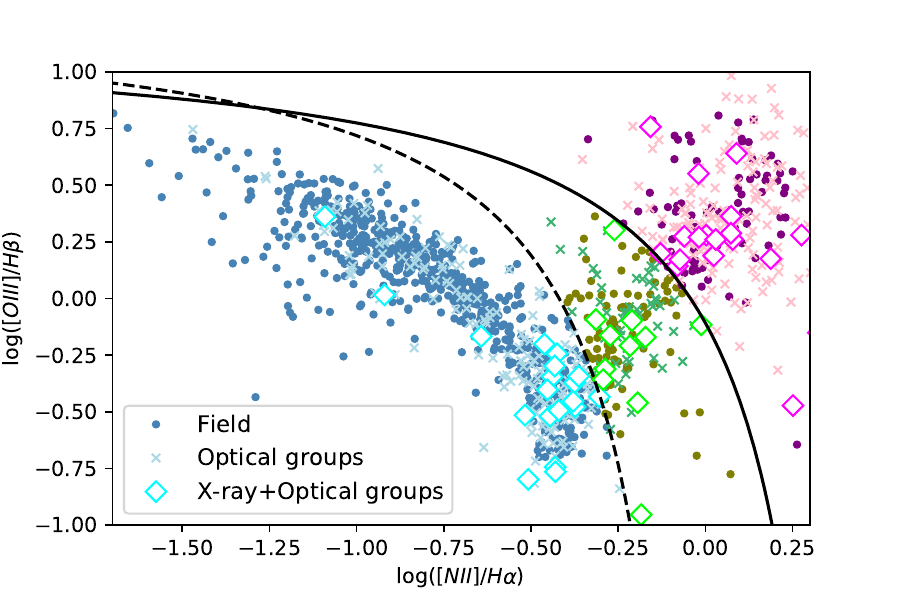}}
    \caption{BPT diagram to select star-forming galaxies (blue-based colours) for the three local environments (circles for Field, crosses for Optical groups and diamonds for X-ray+Optical groups). The dashed and solid curves are the star-forming limit of \citet{Kauffmann2003} and the theoretical maximal star-forming limit of \citet{Kewley2001}, respectively. Green-based colours represent composite galaxies, while pink-based colours are AGN-like galaxies.}
    \label{BPT_SF}
\end{figure} 

To explore the environmental mechanisms causing star formation quenching in the local and large-scale environments, we examine the spatially-resolved distribution of H$\alpha$ emission for star-forming galaxies. These star-forming galaxies are selected according to the Baldwin, Phillips \& Terlevich (BPT) emission-line diagnostics \citep{Baldwin1981,Veilleux1987}. The BPT diagram separates star-forming galaxies from those dominated by an AGN, and composite galaxies, using the H$\alpha$, H$\beta$, [N\,II], and [O\,III] line flux ratios \citep{Kewley2001,Kauffmann2003}. The emission lines are fitted with Gaussian profiles having a common width using the fitting software {\sc lzifu} \citep{Ho2016}. We measure the emission-line fluxes within $1\,R_{\rm e}$, since similar results are obtained by using the mean values of the fluxes per spaxel within $1\,R_{\rm e}$ \citep{Oh2022, Barsanti2023}. A signal-to-noise higher than 3 for each emission line is required. Figure~\ref{BPT_SF} shows the BPT diagram used to select star-forming galaxies for the three local environments. There are 29, 276 and 657 star-forming galaxies respectively within the X-ray+Optical groups, Optical groups, and the Field. 

\begin{figure*} 
    \centering
    {\includegraphics[width=\textwidth]{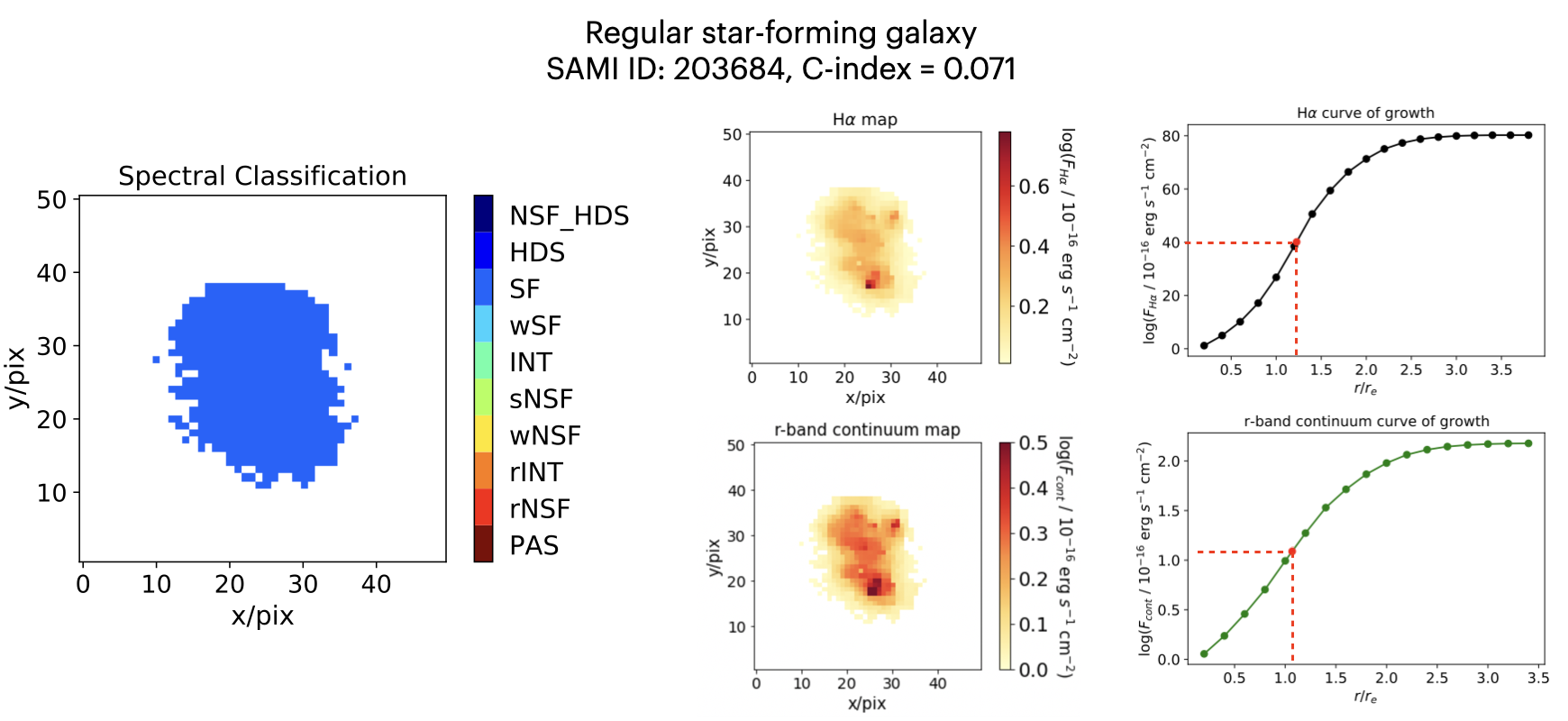}} \\
    \vspace{0.5cm}
    {\includegraphics[width=\textwidth]{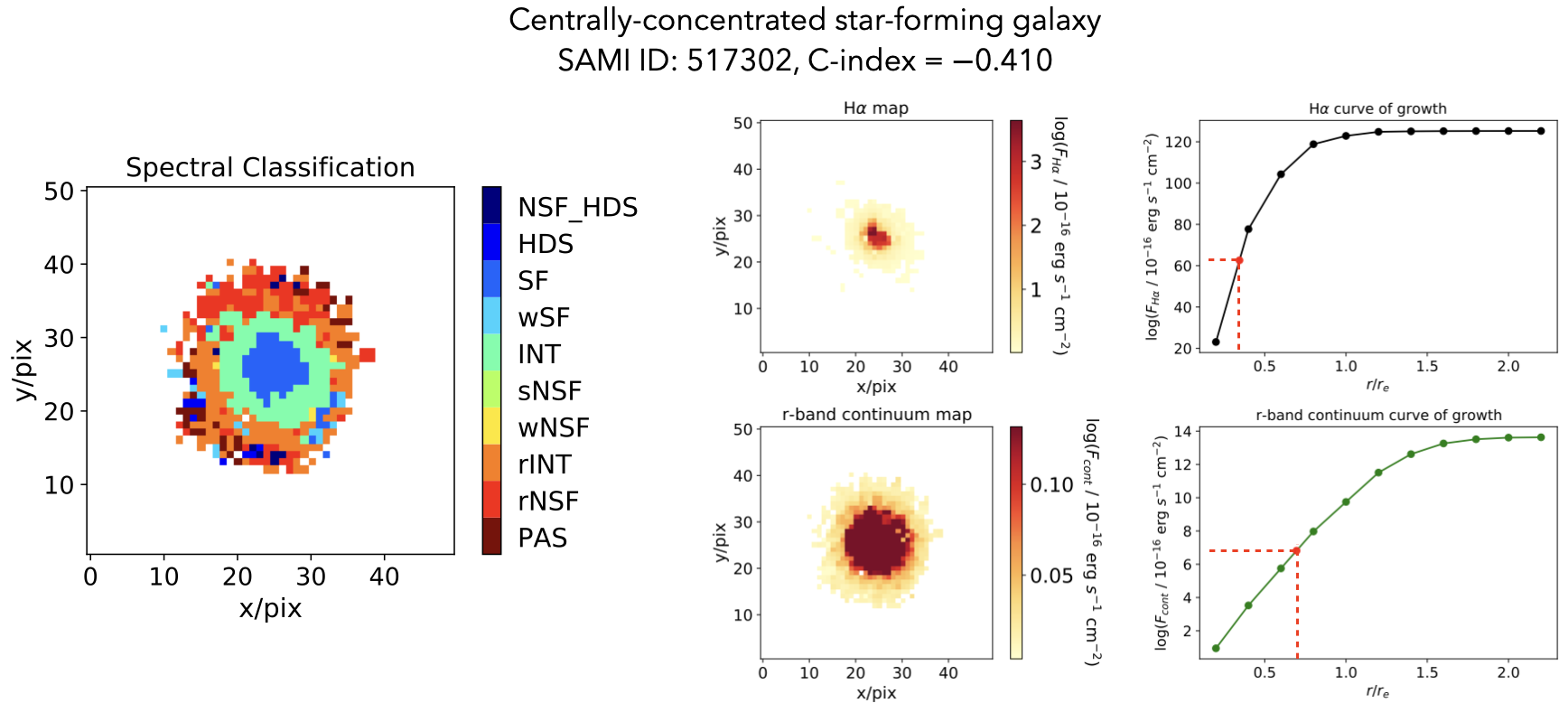}}
    \caption{Examples of a regular star-forming galaxy with C-index$=0.071$ (SAMI ID 203684, top panels) and a centrally-concentrated star-forming galaxy C-index $=-0.410$ (SAMI ID 517302, bottom panels). We show the spatially-resolved spectral classification (left), the H$\alpha$ emission flux map with the associated curve of growth used as tracer for recent star formation (top middle and top right), and the $r$-band continuum flux map with the associated curve of growth used as tracer for the overall stellar distribution (bottom middle and bottom right). The dashed lines mark the galaxy radius enclosing 50\% of the flux. The centrally-concentrated star-forming galaxy is characterised by SF spaxels and high H$\alpha$ flux within the central region.}
    \label{Examples_Cindex}
\end{figure*}

For a more robust selection of star-forming galaxies, we exploit the spatially-resolved spectral classification of \citet{Owers2019}. Each spaxel is given one of ten spectroscopic classifications using a combination of emission and absorption line diagnostics (see their Table~1): passive (PAS), non-star forming (NSF), strong non-star forming (sNSF), weak non-star forming (wNSF), retired non-star forming (rNSF), star forming (SF), weak star forming (wSF), intermediate (INT), retired intermediate (rINT), H$\delta$-strong/post-star forming (HDS), and non-star forming H$\delta$-strong (NSF\_HDS). They only include spaxels with continuum S/N(4100\AA)$\,>\,3$\,pix$^{-1}$ to ensure that the continuum fits are reliable and that both emission- and absorption-line classification is possible. The left panels of Figure~\ref{Examples_Cindex} show examples of the spectral classification for two star-forming galaxies. SAMI galaxy 203684 (top) has all spaxels classified as SF (a regular star-forming galaxy), while SAMI galaxy 517302 (bottom) has SF spaxels concentrated within the central region (a centrally-concentrated star-forming galaxy). 

As defined by \citet{Owers2019}, a galaxy is considered star-forming if at least 10\% of its spaxels are classified as either INT or SF. Based on this criterion, 100\%, 98\%, and 96\% of the BPT-selected star-forming galaxies in the X-ray+Optical groups, Optical groups, and the Field, respectively, meet this requirement. Those galaxies that do not meet this criterion may have limited star-forming regions, low signal-to-noise in their emission lines, or contamination from AGN or shocks that affect the spaxel classification.

C-index measurements are estimated by \citet{Wang2022} and can be used as an indicator for galaxies currently undergoing outside-in star formation quenching. Such measurements are based on comparing the spatially-resolved H$\alpha$ emission flux map to the $r$-band continuum flux map (see Equation~1). For the GAMA equatorial regions, the $r$-band photometry is taken from the SDSS DR 7 imaging \citep{Kelvin2012}. The H$\alpha$ emission flux maps are corrected for 
extinction derived from the Balmer decrement and using the dust law of \citep{Cardelli1989}.

Examples of such maps are shown in Figure~\ref{Examples_Cindex} for a regular star-forming galaxy with C-index $=0.071$ (top panels) and H$\alpha$ emission spread through the whole galaxy and for a centrally-concentrated star-forming galaxy with C-index $=-0.410$ (bottom panels) and H$\alpha$ emission only within the central region. The curves of growth representing the cumulative fluxes measured as a function of radial distance from the galaxy centre are also reported for the H$\alpha$ emission flux and for the $r$-band continuum flux in Figure~\ref{Examples_Cindex}. The dashed lines indicate the radii enclosing 50\% of the flux, $r_{50,{\rm H}\alpha}$ and $r_{50,\rm cont}$, used in the definition of C-index (Equation~\ref{eqn:C-index}). These radii are measured using elliptical isophotes aligned with the galaxy’s projected axis ratio and position angle, rather than circular apertures, providing a closer representation of the intrinsic light distribution in the galaxy plane. This choice minimises inclination-related biases and allows for a more direct comparison of concentration indices across galaxies with different orientations. The H$\alpha$ curve of growth for the centrally-concentrated star-forming galaxy shows a steep rise near the galaxy centre.

To measure C-index, \citet{Wang2022} classified star-forming galaxies into multiple categories using spatially-resolved BPT diagrams for each galaxy spaxel where the emission lines have S/N\,$>$\,2. Galaxies with central spaxels beneath the Kauffmann curve are classified as pure star-forming if they have more than 2/3 of the spaxels below the line, or as central star-forming if they have more than 2/3 of galaxy spaxels in the composite/LINER region. These latter galaxies have star-forming centres and extended LINER/AGN features. Galaxies with central spectrum above the Kauffmann line are classified as central LINER/AGN-like galaxies if they have LINER/AGN centres with extended star formation discs or pure AGN if they have more than 2/3 of the spaxels above the Kewley curve. This classification is used to correct C-index measurements by isolating the H$\alpha$ flux originating from star formation, excluding contributions from harder ionizing sources such as LINER or AGN activity. This correction (see Section~3.2 of \citealt{Wang2022} for technical details) allows inclusion of galaxies with star-forming discs but passive centres that may have a weak central LINER/AGN-like emission.

C-index measurements are available for 649 SAMI star-forming galaxies with $7.5<\log(M_{\star}/M_{\odot})<12$ distributed as follows: 22 star-forming galaxies in X-ray+Optical groups (21 pure star-forming, 1 with LINER/AGN correction), 276 star-forming galaxies in Optical groups (216 pure star-forming, 60 with LINER/AGN correction), and 351 star-forming galaxies in the Field (318 pure star-forming, 33 with LINER/AGN correction). Relative to the large-scale structure, the 649 SAMI star-forming galaxies are distributed as follows: 126 are in Nodes, 211 are Filaments and 312 are in Voids.

\subsection{Galaxy spin--filament alignments}
\label{Spin Flips}

To better understand the physical processes shaping the spatial distribution of star formation, we explore the connection between C-index and the kinematic alignment of galaxy spin with respect to the closest filament. We use measurements of galaxy spin--filament alignments for SAMI galaxies from \citet{Barsanti2022}. Briefly, the galaxy spin--filament alignment is parametrised as the absolute value of the cosine of the angle between the galaxy spin axis and the closest filament in 3D Cartesian coordinates (e.g., \citealp{Tempel2013b,Kraljic2021})
\begin{equation}
|\cos\gamma|=\frac{|\mathbf{L} \cdot \mathbf{r}|}{|\mathbf{L}| \cdot |\mathbf{r}|}
\end{equation}
where \textbf{L} is the galaxy spin axis, determined using spatially-resolved stellar kinematics (which sets the lower stellar mass limit for reliably measuring kinematic position angles) and inclination angles, and \textbf{r} is the orientation vector of the filament. The angle |$\cos\gamma$| varies in the range [0,1], with |$\cos\gamma$|=1 meaning the galaxy spin axis is parallel to the filament while |$\cos\gamma$|=0 means the galaxy spin axis is perpendicular to the filament. Of the 649 SAMI star-forming galaxies with available C-index measurements, there are 375 with measured stellar spin--filament alignments spanning $9.5<\log(M_{\star}/M_{\odot})<12$.

Taking advantage of the SAMI spatially-resolved H$\alpha$ velocity maps, \citet{Barsanti2022} measured also ionised gas spin--filament alignments, where the orientation of the galaxy spin axes are identified using the
gas kinematic position angles. Of the 649 SAMI star-forming galaxies with available C-index measurements, there are 635 with measured ionised gas spin--filament alignments.

\section{Results}
\label{Results}

\begin{figure}
    \centering
    {\includegraphics[width=\columnwidth]{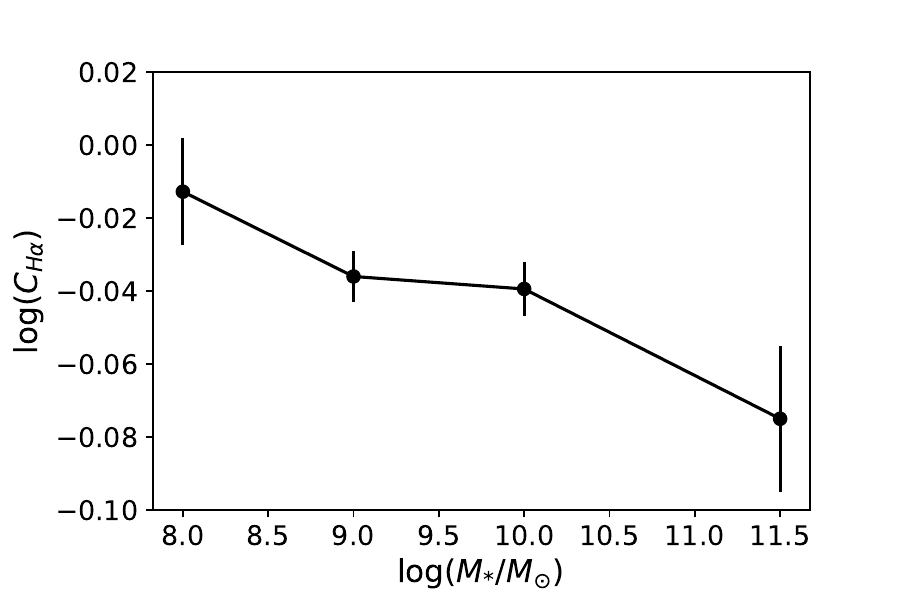}}
    \caption{Mean C-index as a function of stellar mass for 649 SAMI star-forming galaxies. As $M_{\star}$ increases, C-index decreases, meaning that higher-mass galaxies tend to have more centrally-concentrated star formation. }
    \label{Cindex_Mass}
\end{figure}

We aim to explore spatially-resolved star formation quenching in local and large-scale galaxy habitats to understand physical processes regulating galaxy evolution on different environmental scales. Since stellar mass plays a fundamental role in shaping star formation \citep{Peng2010}, we first assess the relation between C-index and stellar mass in Figure~\ref{Cindex_Mass} for 649 SAMI star-forming galaxies. As $M_{\star}$ increases, C-index decreases, meaning that higher-mass galaxies tend to have more centrally-concentrated star formation. The Spearman rank correlation test shows a correlation coefficient $\rho=-0.15$ and p-value $p_S=10^{-4}$.

Across different local environments, we find that the correlation between C-index and stellar mass persists for Field galaxies ($\rho=-0.22$, $p_S=10^{-3}$) and is also present, though weaker, in Optical groups ($\rho=-0.14$, $p_S=0.023$). In contrast, no significant correlation is detected in X-ray+Optical groups ($\rho=-0.09$, $p_S=0.537$). Within the cosmic web, the relation between C-index and stellar mass is found in Voids ($\rho=-0.20$, $p_S=10^{-4}$) and Filaments ($\rho=-0.18$, $p_S=10^{-3}$), while it is not significant for Nodes ($\rho=-0.12$, $p_S=0.101$). This suggests that in high-density environments, external environmental processes dominate over stellar mass in regulating the spatial distribution of star formation.

Similar conclusions are found for the correlations between C-index and bulge-to-total ratio (B/T) in the different environments. The latter is estimated from the 2D photometric bulge/disc decomposition of \citet{Casura2022}. The disc is defined as the exponential component, while the bulge corresponds to the S\'ersic component representing the light excess over the disc. The photometric decomposition uses the image analysis package {\sc ProFound} and the photometric galaxy profile fitting package {\sc ProFit} \citep{Robotham2017,Robotham2018}. For low-dense local and large-scale environments (Field, Voids and Filaments) there is a significant correlation for C-index with B/T ($p_S=0.003$; $p_S=10^{-4}$; $p_S=0.001$, respectively), while for high-dense local and large-scale environments (Optical groups, X-ray+Optical groups, Nodes) there is no correlation ($p_S=0.127$; $p_S=0.406$; $p_S=0.374$, respectively). This confirms that environmental processes play a major role in high-dense environments in shaping the spatial distribution of star formation.

\subsection{Concentrated star formation in local environments}
\label{Concentrated star formation in local environments}

\begin{figure}
    \centering
    {\includegraphics[width=\columnwidth]{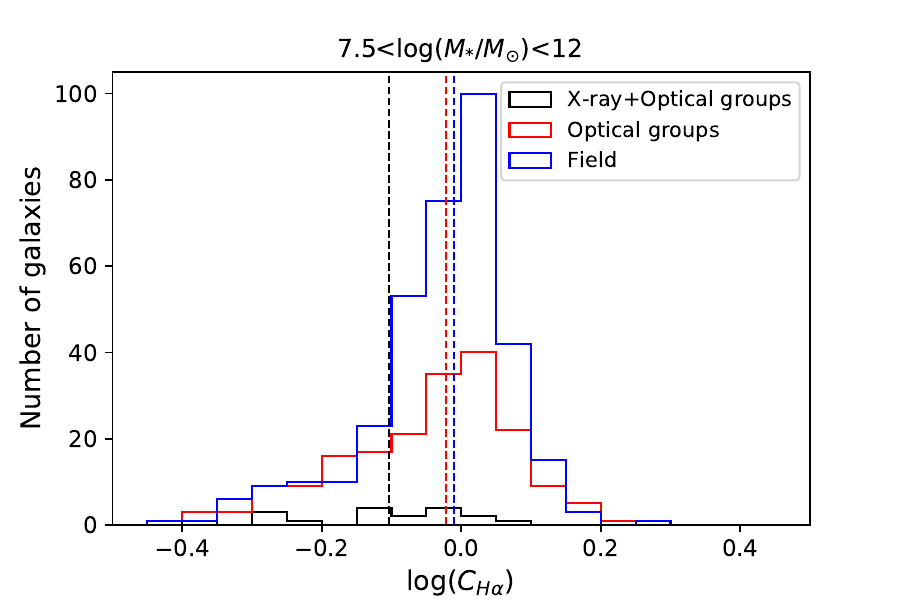}}
    \caption{C-index distributions for star-forming galaxies in X-ray+Optical groups (black), Optical groups (red), and the Field (blue). The median C-index (dashed line) is lowest for X-ray+Optical groups.}
    \label{Cindex_SF}
\end{figure}

We investigate the significance of outside-in star formation quenching in the various local environments in Figure~\ref{Cindex_SF}, which shows C-index distributions for star-forming galaxies in X-ray+Optical groups, Optical groups, and the Field. The median C-index value is lowest for X-ray+Optical groups ($-0.104\pm$0.034, see Table~\ref{Cindex_medians_fractions}) with respect to Optical groups ($-0.022\pm$0.009) and the Field ($-0.010\pm$0.005). X-ray+Optical groups are the local environments hosting the largest fraction of centrally-concentrated star-forming galaxies ($f_{\rm CC}$) with C-index $<-0.2$. This is the same C-index cut used by \citet{Schaefer2016} and \citet{Wang2022} to define a centrally-concentrated star-forming galaxy, and is two standard deviations below the median for the whole star-forming sample. We find $f_{\rm CC}=32^{+12}_{-9}\%$ for X-ray+Optical groups, compared to $f_{\rm CC}=13^{+3}_{-2}\%$ for Optical groups and $f_{\rm CC}=8^{+2}_{-1}\%$ for the Field (see Table~\ref{Cindex_medians_fractions}). The uncertainties on these fractions are calculated using the binomial confidence intervals \citep{Cameron2011}. Since for small samples, such as the X-ray+Optical groups, the fraction of centrally-concentrated star-forming galaxies may be sensitive to the exact placement of the C-index threshold, we verify that these results are robust to modest variations of the threshold. We repeat the analysis using thresholds of C-index $<-0.15$ and C-index $<-0.25$, finding that the environmental trends of $f_{\rm CC}$ remain unchanged within the uncertainties.

Results from the two-sample K-S test reported in Table~\ref{Cindex_KS} reveal statistically significant differences among the three distributions. Due to the small number of star-forming galaxies (22) for the X-ray+Optical groups compared to the Optical groups (276) and the Field (351), we  perform a two-sample KS test with 1000 bootstrap samples to estimate the significance robustly. 

Moreover, given the highly imbalanced sample sizes between environments, particularly for X-ray+Optical groups, we complement the K–S test with a non-parametric effect-size measure, Cliff’s delta, which quantifies the degree of stochastic dominance between two distributions and is less sensitive to sample-size imbalance \citep{Meissel2024}. Cliff’s delta is a non-parametric effect-size measure closely related to the Mann–Whitney U test \citep{MannWhitney1947}. Unlike a pure significance test, it is both signed, indicating the direction of the difference between two distributions, and normalised, allowing a direct and intuitive comparison of effect strengths across different samples. In Table~\ref{Cindex_KS} we also report Cliff’s $\delta$ together with 95\% bootstrap confidence intervals (C.I.) to assess the magnitude and robustness of differences in the C-index distributions. Negative $\delta$ values indicate lower C-index, i.e. more centrally-concentrated star formation, in Sample~1 relative to Sample~2. The 95\% C.I. are estimated from 1000 bootstrap realisations. Repeating the bootstrap procedure 50 times yields consistent results, indicating that the inferred C.I. are stable despite the limited sample size for the X-ray+Optical group galaxies. The corresponding effect-size estimates confirm a systematic modest shift toward lower C-index values in X-ray+Optical groups relative to Optical groups and the Field. We further discuss the limitations of small galaxy samples in Section~\ref{Limitations due to small galaxy samples}.

\begin{figure*}
    \centering
    {\includegraphics[width=0.49\textwidth]{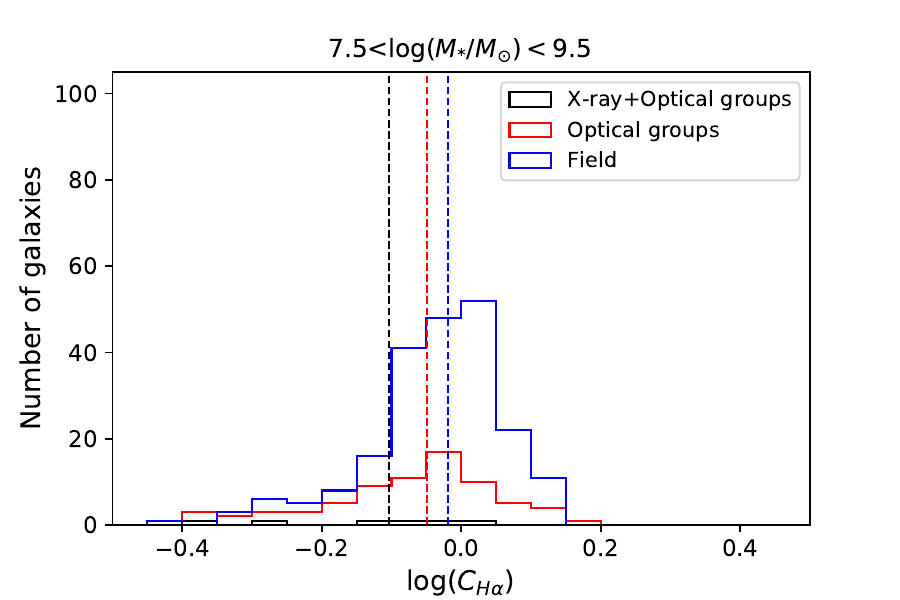}} 
    {\includegraphics[width=0.49\textwidth]{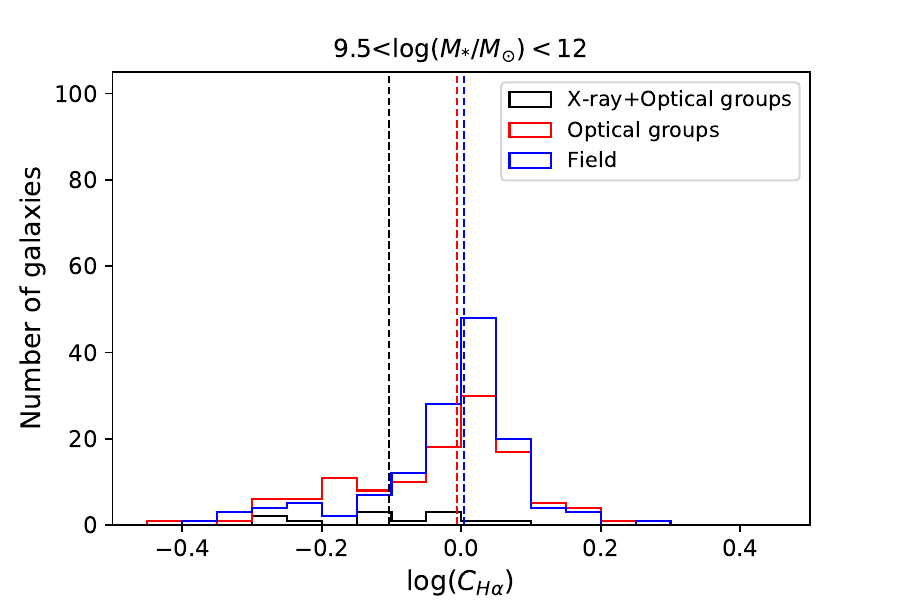}}
    \caption{C-index distributions for star-forming galaxies with $7.5<\log(M_{\star}/M_{\odot})<9.5$ (left panel) and $9.5<\log(M_{\star}/M_{\odot})<12$ (right panel) in local environments. The median C-index is lowest for X-ray+Optical groups.}
    \label{Cindex_SF_stellarMass}
\end{figure*}

\begin{figure*}
    \centering
    {\includegraphics[width=0.49\textwidth]{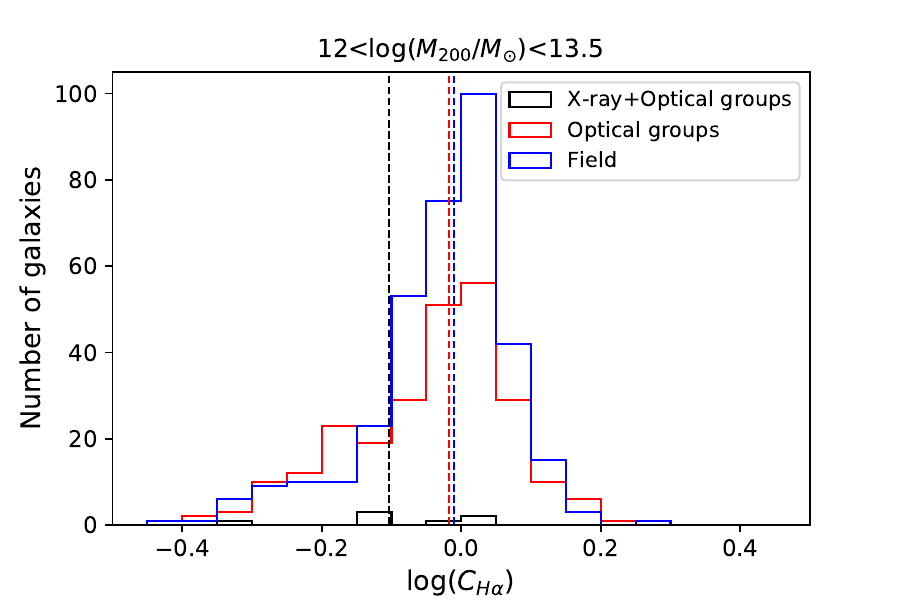}} 
    {\includegraphics[width=0.49\textwidth]{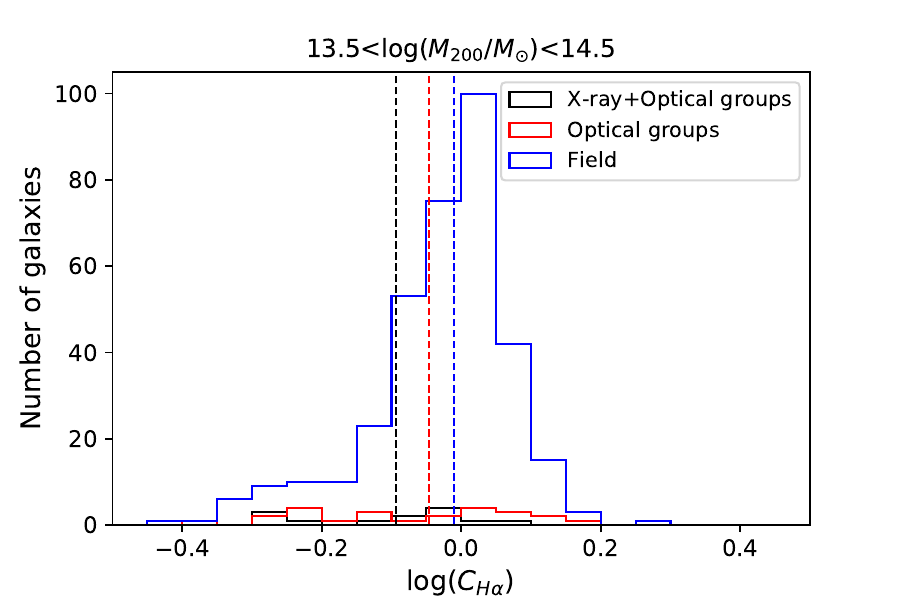}}
    \caption{C-index distributions for star-forming galaxies in X-ray+Optical groups and Optical groups with $12<\log(M_{200}/M_{\odot})<13.5$ (left panel) and with $13.5<\log(M_{200}/M_{\odot})<14.5$ (right panel), and for the whole Field sample. The median C-index  is lowest for X-ray+Optical groups.}
    \label{Cindex_SF_haloMass}
\end{figure*}

To understand the impact of stellar mass, we divide the star-forming sample into low-mass galaxies with $7.5<\log(M_{\star}/M_{\odot})<9.5$ and high-mass galaxies with $9.5<\log(M_{\star}/M_{\odot})<12$. X-ray+Optical groups still have the lowest median C-index and the largest fraction of centrally-concentrated star-forming galaxies compared to Optical groups and the Field in both stellar mass ranges (see Figure~\ref{Cindex_SF_stellarMass} and Tables~\ref{Cindex_medians_fractions} \&~\ref{Cindex_KS}). To explore the role of halo mass, we separate low-mass halos with $12<\log(M_{200}/M_{\odot})<13.5$ from high-mass halos with $13.5<\log(M_{200}/M_{\odot})<14.5$ for both X-ray+Optical groups and Optical groups. For the high-mass halos, there is no significant difference between the $M_{200}$ distributions for these local environments (see Figure~\ref{M200_Mstar_histograms}). For both halo mass ranges, the lowest C-index median and the largest fraction of centrally-concentrated star-forming galaxies is found for X-ray+Optical groups (see Figure~\ref{Cindex_SF_haloMass} and Tables~\ref{Cindex_medians_fractions} \&~\ref{Cindex_KS}). The corresponding Cliff’s delta effect-size estimates further support these trends, indicating a systematic moderate shift toward lower C-index values in X-ray+Optical groups relative to Optical groups and the Field across both stellar mass and halo mass bins, albeit with substantial overlap between the distributions due to the limited sample size.

We perform stellar mass-matched analysis of the local environmental trends for the C-index distributions in Appendix~\ref{Stellar mass-matched analysis of environmental trends}. The same results are found, confirming that the enhanced central concentration of star formation in X-ray+Optical groups is not driven by stellar mass differences.

These results suggest a trend toward enhanced outside-in quenching and more centrally concentrated star formation in X-ray+Optical groups compared to Optical groups and the Field, regardless of stellar mass or group mass, although the systematic shift towards lower C-index values are modest.

\begin{table*}
\caption{Median C-index for star-forming galaxies in different local environments (X-ray+Optical groups, Optical groups, and the Field). Column~1 lists the environment, column~2 the $M_{\star}$ range, column~3 the $M_{200}$ range, column~4 reports the median C-index (with bootstrap uncertainty), and column~5 lists the fraction of centrally-concentrated star-forming galaxies with C-index $<-0.2$, with binomial confidence intervals \citep{Cameron2011}.}
\centering
\begin{tabular}{lcccc}
\toprule
Local Environment & $M_{\star}$ range & $M_{200}$ range & $\langle$C-index$\rangle$ & $f_{\rm CC}$ (\%) \\
\midrule
X-ray+Optical Groups & $\left[7.5; 12.0\right]$ & $\left[12; 14.5\right]$ & $-0.104\pm0.034$ & $32^{+12}_{-9}$ \\
Optical Groups & $\left[7.5; 12.0\right]$ & $\left[12; 14.5\right]$ & $-0.022\pm0.009$ & $13^{+3}_{-2}$ \\
Field & $\left[7.5; 12.0\right]$ & -- & $-0.010\pm0.005$ & $8^{+2}_{-1}$ \\
\midrule
X-ray+Optical Groups & $\left[7.5; 9.5\right]$ & $\left[12; 14.5\right]$ & $-0.104\pm0.071$ & $43^{+19}_{-15}$ \\
Optical Groups & $\left[7.5; 9.5\right]$ & $\left[12; 14.5\right]$ & $-0.049\pm0.014$ & $15^{+5}_{-4}$ \\
Field & $\left[7.5; 9.5\right]$ & -- & $-0.021\pm0.007$ & $7^{+3}_{-2}$ \\
\midrule
X-ray+Optical Groups & $\left[9.5; 12.0\right]$ & $\left[12; 14.5\right]$ & $-0.104\pm0.030$ & $25^{+6}_{-5}$ \\
Optical Groups & $\left[9.5; 12.0\right]$ & $\left[12; 14.5\right]$ & $-0.007\pm0.011$ & $12^{+4}_{-3}$ \\
Field & $\left[9.5; 12.0\right]$ & -- & $+0.004\pm0.009$ & $9^{+3}_{-2}$ \\
\midrule
X-ray+Optical Groups & $\left[7.5; 12.0\right]$ & $\left[12; 13.5\right]$ & $-0.104\pm0.044$ & $14^{+6}_{-4}$ \\
Optical Groups & $\left[7.5; 12.0\right]$ & $\left[12; 13.5\right]$ & $-0.018\pm0.007$ & $11^{+3}_{-2}$ \\
\midrule
X-ray+Optical Groups & $\left[7.5; 12.0\right]$ & $\left[13.5; 14.5\right]$ & $-0.093\pm0.041$ & $33^{+9}_{-7}$ \\
Optical Groups & $\left[7.5; 12.0\right]$ & $\left[13.5; 14.5\right]$ & $-0.046\pm0.031$ & $29^{+7}_{-6}$ \\
\bottomrule
\end{tabular}
\label{Cindex_medians_fractions}
\end{table*}

\begin{table*}
\caption{Results from two-sample Kolmogorov-Smirnov test with bootstrap resampling and non-parametric effect-size estimates for the C-index distributions of star-forming galaxies in local environments (X-ray+Optical groups, Optical groups and Field). Columns~1 and~2 list the $M_{\star}$ and the $M_{200}$ ranges, columns~3 and~4 give the two samples being compared (with numbers of star-forming galaxies in [..]); the associated $p$-value is shown in column~5, with significant $p$-values ($<$0.05) in bold. Column~6 lists Cliff’s delta $\delta$ (negative values indicate lower C-index, i.e. more centrally-concentrated star formation, in Sample~1 relative to Sample~2), with 95\% bootstrap confidence intervals.}
\centering
\begin{tabular}{ccllcc}
\toprule
$M_{\star}$ range & $M_{200}$ range & Sample 1 [N$_{\rm gal}$] & Sample 2 [N$_{\rm gal}$] & $p_{\rm 2KS}$ & Cliff’s $\delta$ (95\% C.I.)\\
 \midrule
$\left[7.5; 12.0\right]$ & $\left[12; 14.5\right]$& X-ray+Optical groups [22] & Optical groups [276] & \textbf{0.024} & $-0.365\,(-0.590, -0.124)$\\
$\left[7.5; 12.0\right]$&$\left[12; 14.5\right]$& X-ray+Optical groups [22] & Field [351] & \textbf{0.001} & $-0.491\,(-0.702, -0.252)$\\
$\left[7.5; 12.0\right]$ & $\left[12; 14.5\right]$& Optical groups [276] & Field [351] & \textbf{0.011} & $-0.092\,(-0.194, +0.014)$\\
\midrule
$\left[7.5; 9.5\right]$ & $\left[12; 14.5\right]$& X-ray+Optical groups [8] & Optical groups [115] & \textbf{0.023} & $-0.409\,(-0.789, +0.022)$\\
$\left[7.5; 9.5\right]$&$\left[12; 14.5\right]$& X-ray+Optical groups [8] & Field [215] & \textbf{0.013} & $-0.581\,(-0.875, -0.229)$\\
$\left[7.5; 9.5\right]$ & $\left[12; 14.5\right]$& Optical groups [115] & Field [215] & \textbf{0.041} & $-0.173\,(-0.328, -0.014)$ \\
\midrule
$\left[9.5; 12.0\right]$ & $\left[12; 14.5\right]$& X-ray+Optical groups [14] & Optical groups [161] & \textbf{0.043} & $-0.343\,(-0.619, -0.042)$\\
$\left[9.5; 12.0\right]$&$\left[12; 14.5\right]$& X-ray+Optical groups [14] & Field [136] & \textbf{0.008} & $-0.480\,(-0.744, -0.173)$\\
$\left[9.5; 12.0\right]$ & $\left[12; 14.5\right]$& Optical groups [161] & Field [136] & 0.116 & $-0.081\,(-0.226, +0.062)$\\
\midrule
$\left[7.5; 12.0\right]$ & $\left[12; 13.5\right]$& X-ray+Optical groups [7] & Optical groups [252] & \textbf{0.036} & $-0.241\,(-0.596, -0.130)$\\
$\left[7.5; 12.0\right]$&$\left[12; 13.5\right]$& X-ray+Optical groups [7] & Field [351] & \textbf{0.025} & $-0.372\,(-0.732, -0.014)$\\
$\left[7.5; 12.0\right]$ & $\left[12; 13.5\right]$& Optical groups [252] & Field [351] & 0.135 & $-0.073\,(-0.167, +0.022)$\\
\midrule
$\left[7.5; 12.0\right]$ & $\left[13.5; 14.5\right]$& X-ray+Optical groups [15] & Optical groups [24] & \textbf{0.023} & $-0.300\,(-0.633, +0.067]$\\
$\left[7.5; 12.0\right]$&$\left[13.5; 14.5\right]$& X-ray+Optical groups [15] & Field [351] & \textbf{0.011} & $-0.504\,(-0.737, -0.233)$\\
$\left[7.5; 12.0\right]$ & $\left[13.5; 14.5\right]$& Optical groups [24] & Field [351] & \textbf{0.029} & $-0.136\,(-0.427, +0.169)$\\
\bottomrule
\end{tabular}
\label{Cindex_KS}
\end{table*}

\subsection{Concentrated star formation in the cosmic web}
\label{Concentrated star formation in the cosmic web}

\begin{figure}
    \centering
    {\includegraphics[width=\columnwidth]{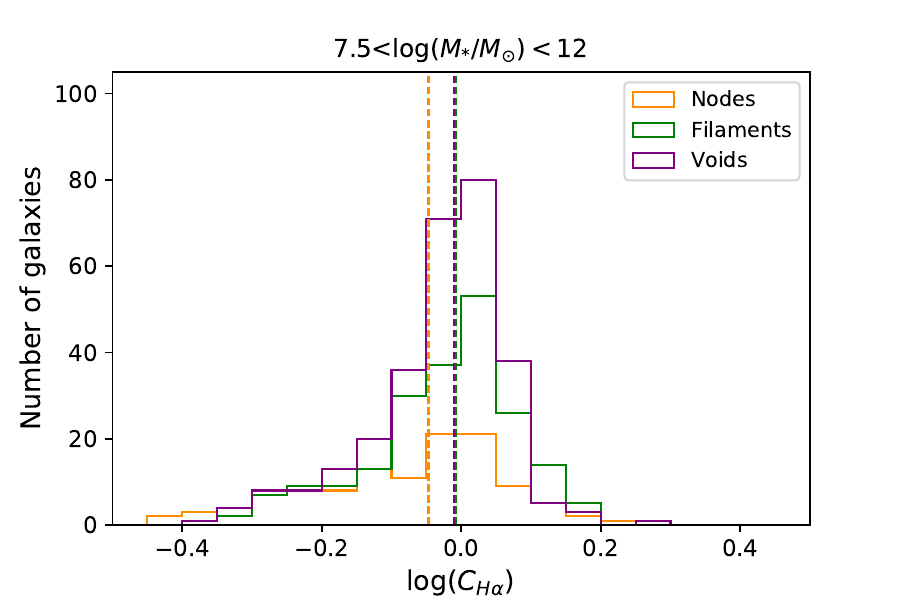}}
    \caption{C-index distributions for star-forming galaxies in Nodes (orange), Filaments (green) and Voids (purple). The median C-index (dashed line) is lowest for Nodes.}
    \label{Cindex_SF_CW}
\end{figure}

We explore the importance of outside-in quenching in the different large-scale environments in Figure~\ref{Cindex_SF_CW}, which compares C-index distributions for star-forming galaxies in Nodes, Filaments and Voids. The median C-index is lowest for Nodes relative to Filaments and Voids (see Table~\ref{Cindex_medians_fractions_CW}). Consistently, the highest fraction of centrally-concentrated star-forming galaxies with C-index $<-0.2$ is found for Nodes ($f_{\rm CC}=21^{+5}_{-4}\%$); Filaments have $f_{\rm CC}=9^{+3}_{-2}\%$ and Voids have $f_{\rm CC}=7^{+3}_{-2}\%$. Nodes exhibit significantly different C-index distributions compared to both Filaments and Voids (Table~\ref{Cindex_KS_CW}), with small systematic shifts toward lower C-index values quantified by Cliff’s delta. In contrast, Filaments and Voids show substantial overlap in their C-index distributions, with no statistically significant difference under either the K–S test or Cliff’s delta.

\begin{figure*}
    \centering
    {\includegraphics[width=0.49\textwidth]{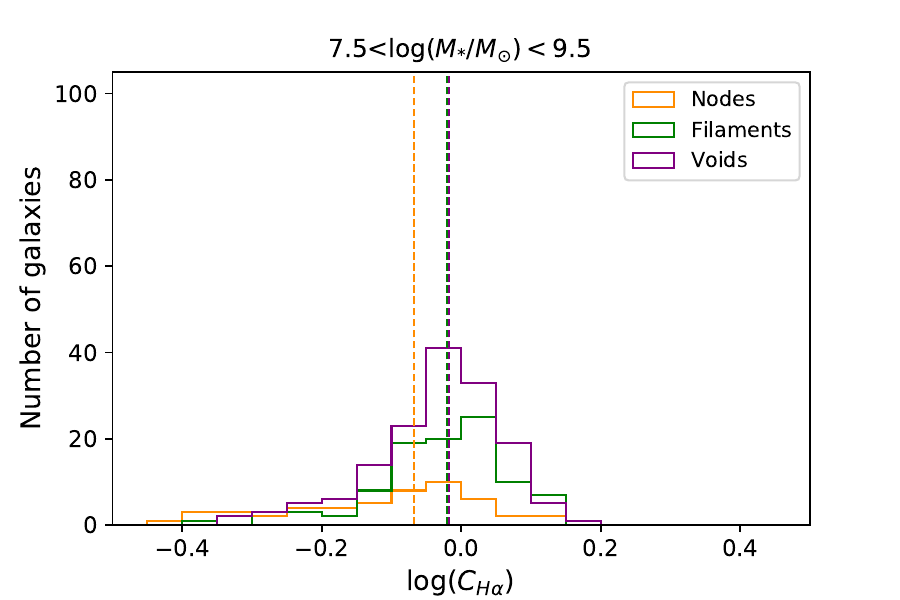}} 
    {\includegraphics[width=0.49\textwidth]{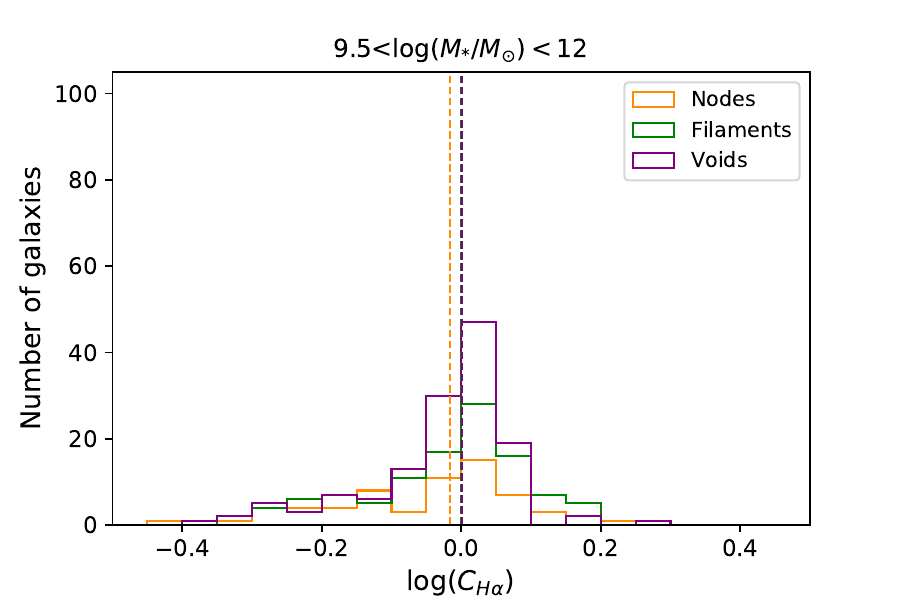}}
    \caption{C-index distributions for star-forming galaxies with $7.5<\log(M_{\star}/M_{\odot})<9.5$ (left panel) and $9.5<\log(M_{\star}/M_{\odot})<12$ (right panel) in large-scale environments. The median C-index lines for voids and filaments overlap, while the median C-index is lowest for Nodes. The difference is significant for low-mass galaxies. }
    \label{Cindex_SF_CW_Mass}
\end{figure*}

The star-forming sample is divided into low-mass galaxies with $7.5<\log(M_{\star}/M_{\odot})<9.5$ and high-mass galaxies with $9.5<\log(M_{\star}/M_{\odot})<12$ in Figure~\ref{Cindex_SF_CW_Mass} and Tables~\ref{Cindex_medians_fractions_CW} \&~\ref{Cindex_KS_CW}. The median C-index is lowest for galaxies in Nodes, and this difference is statistically significant for low-mass star-forming galaxies. This is supported by small-to-moderate Cliff’s delta effect sizes, which indicate a systematic shift toward lower C-index values for Nodes relative to Filaments and Voids, while Filaments and Voids show substantial overlap. In contrast, high-mass star-forming galaxies exhibit no significant differences in C-index across large-scale environments, with Cliff’s delta values close to zero and confidence intervals spanning zero.

Appendix~\ref{Stellar mass-matched analysis of environmental trends} reports the stellar mass-matched analysis for the large-scale environmental trends of the C-index distributions. We find consistent results, indicating that these large-scale environmental trends are not a by-product of stellar mass segregation.

Stronger signs of outside-in quenching are found in Nodes compared to Filaments and Voids, with the trend driven by low-mass galaxies, suggesting that their star formation is more affected by cosmic environment than that of high-mass galaxies.

\begin{table}
\caption{Median C-index for star-forming galaxies in different large-scale environments (nodes, filaments, and voids). Column~1 lists the cosmic environment, column~2 the $M_{\star}$ range, column~3 the median C-index (with bootstrap uncertainty), and column~4 the fraction of centrally-concentrated star-forming galaxies with C-index $<-0.2$, with binomial confidence intervals \citep{Cameron2011}.}
\centering
\begin{tabular}{lccc}
\toprule
Environment & $M_{\star}$ range & $\langle$C-index$\rangle$ & $f_{\rm CC}$ (\%)\\
\midrule
Nodes & $\left[7.5; 12.0\right]$ & $-0.047\pm0.013$ & $21^{+5}_{-4}$\\
Filaments & $\left[7.5; 12.0\right]$ & $-0.011\pm0.006$ & $9^{+3}_{-2}$\\
Voids & $\left[7.5; 12.0\right]$ & $-0.006\pm0.007$ & $7^{+3}_{-2}$\\
\midrule
Nodes & $\left[7.5; 9.5\right]$ & $-0.067\pm0.021$ & $25^{+7}_{-6}$\\
Filaments & $\left[7.5; 9.5\right]$ & $-0.019\pm0.009$ & $7^{+3}_{-2}$\\
Voids & $\left[7.5; 9.5\right]$ & $-0.018\pm0.007$ & $6^{+4}_{-3}$\\
\midrule
Nodes & $\left[9.5; 12.0\right]$ & $-0.017\pm0.017$ & $17^{+6}_{-5}$\\
Filaments & $\left[9.5; 12.0\right]$ & $0.001\pm0.011$ & $11^{+4}_{-3}$\\
Voids & $\left[9.5; 12.0\right]$ & $0.001\pm0.009$ & $8^{+3}_{-2}$\\
\bottomrule
\end{tabular}
\label{Cindex_medians_fractions_CW}
\end{table}

\begin{table*}
\caption{Results from two-sample Kolmogorov-Smirnov test and non-parametric effect-size estimates for the C-index distributions of star-forming galaxies in large-scale environments (Nodes, Filaments, Voids). Column~1 lists the $M_{\star}$ range, columns~2 \&~3 the samples compared, with numbers of star-forming galaxies in [..]; and column~4 the associated $p$-value, with significant $p$-values ($<$0.05) in bold. Column~6 lists Cliff’s delta $\delta$ (negative values indicate lower C-index, i.e. more centrally-concentrated star formation, in Sample~1 relative to Sample~2), with 95\% bootstrap confidence intervals.}
\centering
\begin{tabular}{ccccc}
\toprule
$M_{\star}$ range& Sample 1 [N$_{\rm gal}$] & Sample 2 [N$_{\rm gal}$] & $p_{\rm 2KS}$ & Cliff’s $\delta$ (95\% C.I.) \\
 \midrule
$\left[7.5; 12.0\right]$  & Nodes [126] & Filaments [211] & \textbf{0.001} & $-0.216\,(-0.345, -0.092)$\\
$\left[7.5; 12.0\right]$  & Nodes [126] & Voids [312] & \textbf{0.001} & $-0.191\,(-0.315, -0.065)$\\
$\left[7.5; 12.0\right]$  & Filaments [211] & Voids [312] & 0.192 & $+0.041\,(-0.060, +0.142)$\\
 \midrule
$\left[7.5; 9.5\right]$  & Nodes [60] & Filaments [103] & \textbf{0.002} & $-0.330\,(-0.499, -0.153)$\\
$\left[7.5; 9.5\right]$  & Nodes [60] & Voids [174] & \textbf{0.006} & $-0.281\,(-0.444, -0.114)$\\
$\left[7.5; 9.5\right]$  & Filaments [103] & Voids [174] & 0.843 & $+0.052\,(-0.086, +0.195)$\\
\midrule
$\left[9.5; 12.0\right]$  & Nodes [66] & Filaments [108] & 0.378 & $ -0.117\,(-0.295, +0.064)$\\
$\left[9.5; 12.0\right]$  & Nodes [66] & Voids [138] & 0.063 & $-0.106\,(-0.282, +0.072)$\\
$\left[9.5; 12.0\right]$  & Filaments [108] & Voids [138] & 0.248 & $+0.033\,(-0.115, +0.185)$\\
\bottomrule
\end{tabular}
\label{Cindex_KS_CW}
\end{table*}

\subsection{Correlations of C-index with local and large-scale environment metrics}
\label{Correlations}

\begin{figure*}
    \centering
    {\includegraphics[width=0.45\textwidth]{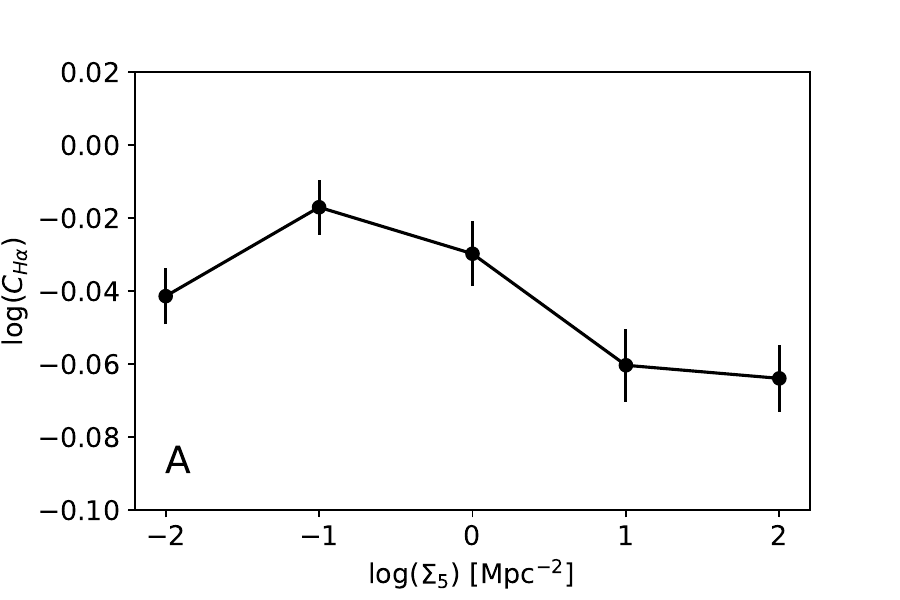}} 
    {\includegraphics[width=0.45\textwidth]{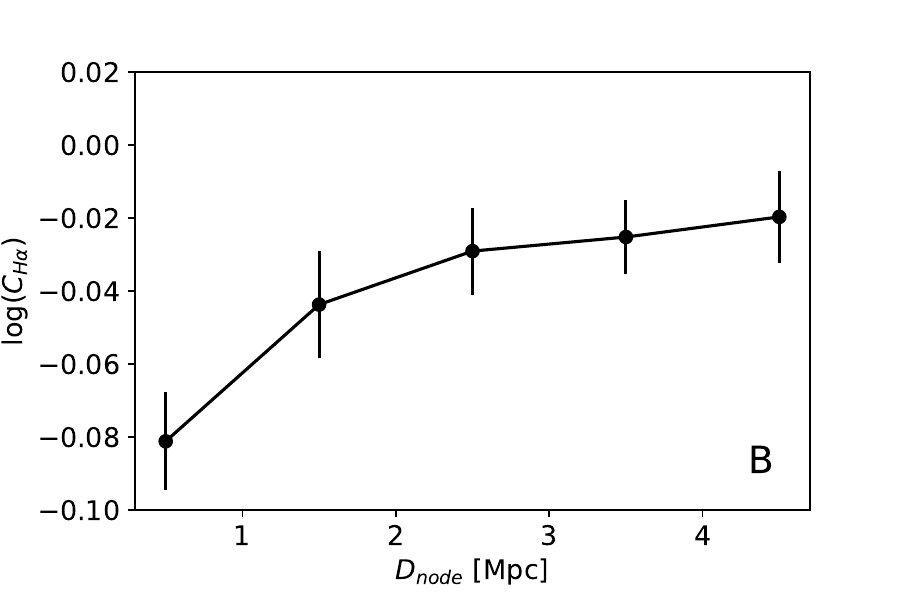}}
    {\includegraphics[width=0.45\textwidth]{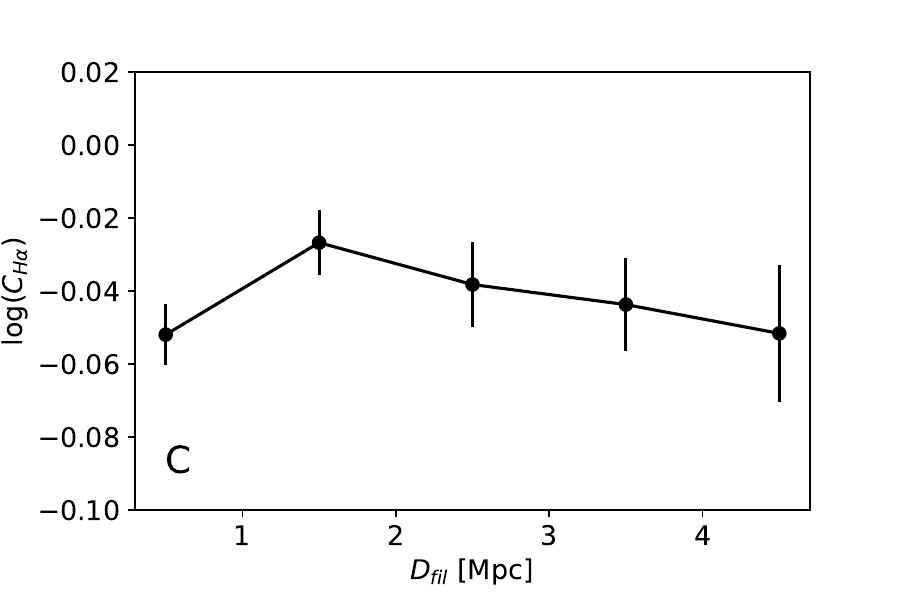}} 
    {\includegraphics[width=0.45\textwidth]{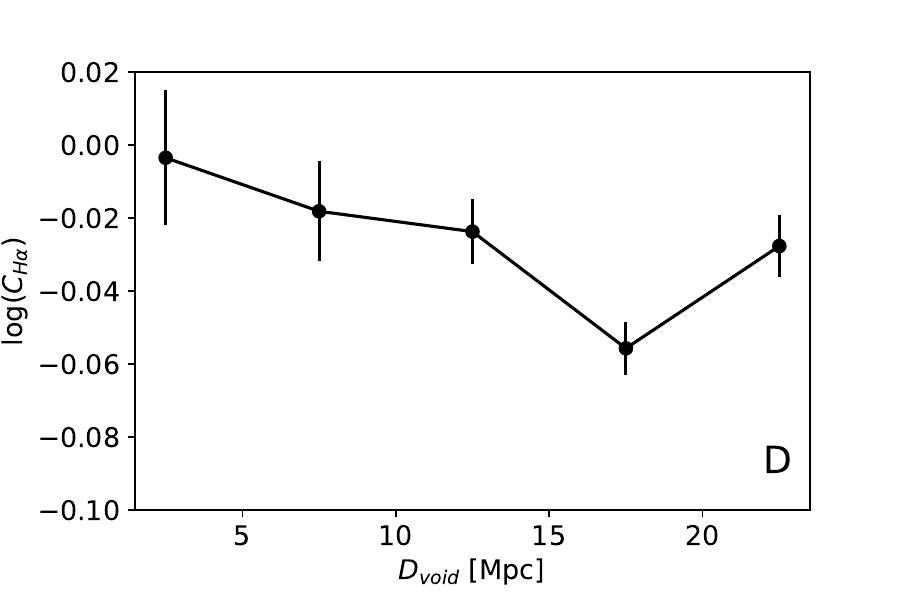}}
    \caption{Mean C-index in bins of local galaxy density (A), distance to node (B), distance to filament (C), and distance to void (D). A lower C-index indicates more centrally concentrated star formation. Standard errors on the mean are shown. The mean C-index decreases with increasing $\Sigma_5$ and $D_{\rm void}$, and with decreasing $D_{\rm node}$; there is no significant change in C-index with $D_{\rm fil}$.}
    \label{Cindex_Env}
\end{figure*} 

To assess the relative importance of different environmental scales on spatially-resolved star formation distribution, we now examine how the C-index varies with the immediate surroundings of each galaxy and its large-scale environmental conditions. We exploit local galaxy density measured as the fifth-nearest neighbour surface density $\Sigma_5=5/\pi d_{5}^{2}$ \citep{Brough2013,Croom2021}. Figure~\ref{Cindex_Env} shows mean C-index as a function of local galaxy density (panel~A), distance to node (panel~B), distance to filament (panel~C), and distance to void (panel~D). Results from the Spearman rank correlation test are reported in Table~\ref{SpearmanResults}. The mean C-index decreases with increasing $\Sigma_{5}$ and $D_{\rm void}$, increases for higher values of $D_{\rm node}$, and shows no trend with $D_{\rm fil}$. For galaxies in groups we tested whether the C-index correlates with halo mass, finding no statistically significant trend. The most significant correlation is measured with $D_{\rm node}$.

\begin{figure*}
\centering
\includegraphics[width=0.73\textwidth]{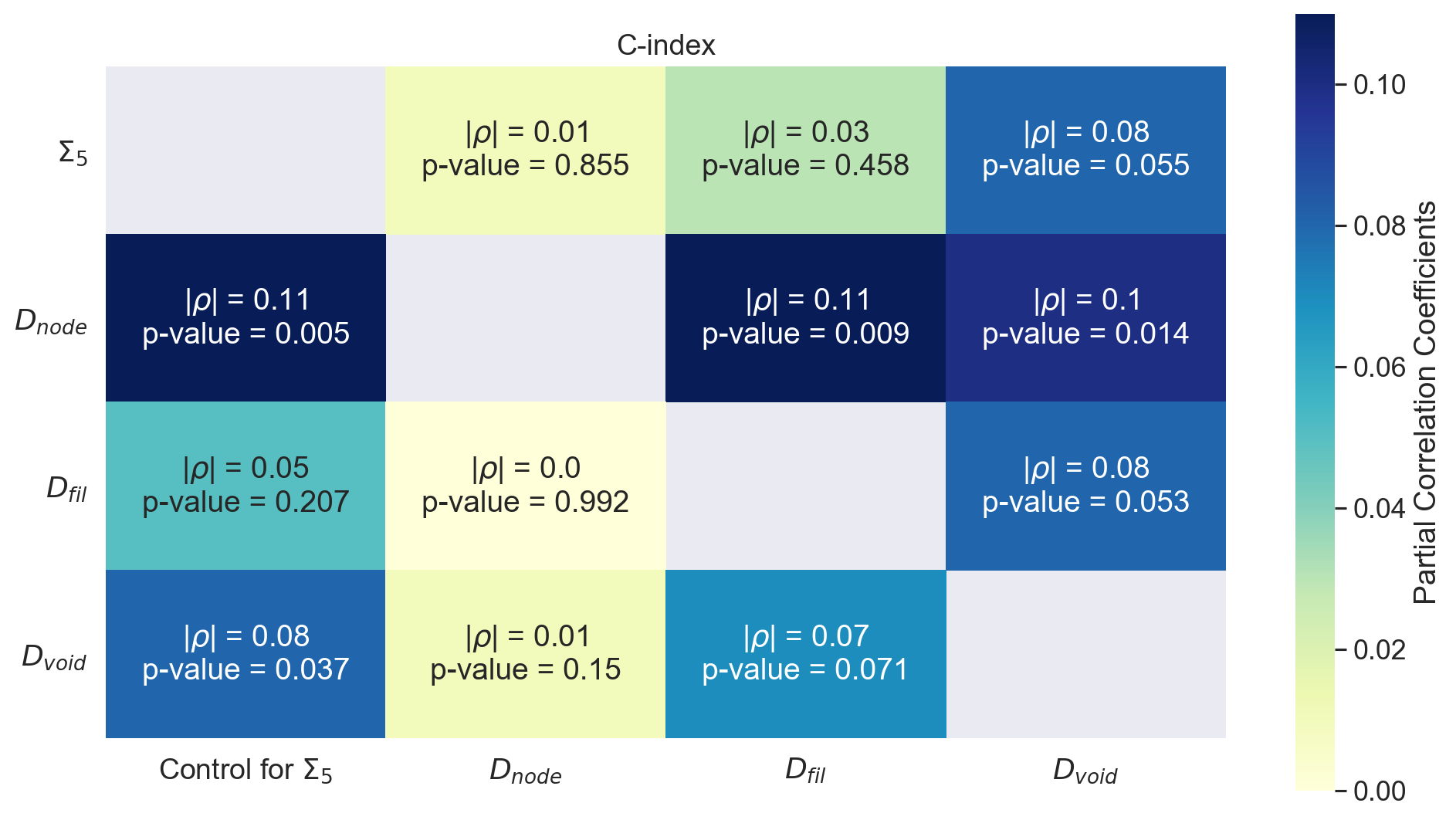}
\caption{Partial correlation coefficients from the Spearman test between C-index and environment metrics while controlling separately for $\Sigma_5$, $D_{\rm node}$, $D_{\rm fil}$ and $D_{\rm void}$. No significant statistical residual correlation is left for any environment metrics while controlling for $D_{\rm node}$.}
\label{PartialCorrelations}
\end{figure*}

\begin{table}
\centering
\caption{Spearman rank correlation tests for C-index as a function of local environment and cosmic web metrics. Column~1 gives the environment metric, column~2 the number of star-forming galaxies in the sample, column~3 the Spearman correlation coefficient, and column~4 the Spearman test $p$-value (bold if $<$0.05).}
\label{SpearmanResults}
\begin{tabular}{@{}lccc@{}}
\toprule
Environment metric &  $N_{\rm gal}$ & $\rho$  & $p_{\rm S}$  \\
\toprule
$\Sigma_{5}$ & $610$ & $-0.07$ & $\textbf{0.029}$\\
$M_{200}$ & $298$ & $-0.11$ & $0.119$\\
\midrule
$D_{\rm node}$ & $610$ & $+0.19$ & \textbf{0.008}\\
$D_{\rm fil}$&  $610$ &$+0.08$ & $0.093$\\
$D_{\rm void}$& $610$ & $-0.11$ & \textbf{0.018}\\
\midrule
$L_{500}$ & $15$ & $-0.57$ & $\textbf{0.037}$ \\
\bottomrule
\end{tabular}
\end{table}

We estimate the C-index partial correlation coefficients to explore the true correlation between two parameters while controlling for a third \citep[avoiding cross-correlation driven by co-dependency on the latter;][]{Lawrance1976}. As Figure~\ref{PartialCorrelations} shows, controlling for $\Sigma_5$ leaves a strong correlation between C-index and $D_{\rm node}$ ($p=0.005$); it is weaker for $D_{\rm void}$ ($p=0.037$) and insignificant for $D_{\rm fil}$ ($p=0.207$). Controlling for $D_{\rm fil}$ or $D_{\rm void}$, the only significant residual correlations are with $D_{\rm node}$ ($p=0.009$ and $p=0.014$, respectively). No significant residual correlation for any other environment metrics remain when controlling for $D_{\rm node}$.

These results imply that proximity to nodes is the key factor in regulating the spatial distribution of star formation in star-forming galaxies, while other environment metrics like $\Sigma_5$, $D_{\rm fil}$, and $D_{\rm void}$ are either subdominant or redundant relative to $D_{\rm node}$.

\subsection{Correlation of C-index with X-ray luminosity}
\label{Correlation of C-index with X-ray luminosity}

Given the dominant significance found for the distance to the closest node in shaping the star formation concentration, we now examine the role of the thermodynamical state of the group environment, traced by X-ray luminosity. X-ray+Optical groups are found closest to nodes (see Figure~\ref{LocalEnvironments_CW}) and show the lowest median C-index and the highest fraction of centrally-concentrated star-forming galaxies (see Tables~\ref{Cindex_medians_fractions} \&~\ref{Cindex_medians_fractions_CW}) out of all the local (X-ray+Optical groups, Optical groups, Field) and large-scale (Nodes, Filaments, Voids) environments. Since the degree of gas depletion and star formation suppression correlates with the X-ray luminosity of the group, suggesting a stronger impact of outside-in quenching in groups with a denser and hotter IGM \citep{Rasmussen2006}, we investigate C-index as a function of X-ray luminosity for 15 star-forming galaxies in 10 X-ray+Optical groups with $13.5<\log(M_{200}/M_{\odot})<14.5$. Specifically, we use $\rm L_{500}$: the soft X-ray band (0.5--2\,keV) integrated out to $\rm R_{500}$ (the radius within which the mean density is 500 times the critical density). We verified that the dynamical halo masses $M_{200}$ of the X-ray+Optical groups show no significant trend with $\rm L_{500}$ ($\rho = 0.40$, $p_S = 0.6$), indicating that any observed trend in central star formation with X-ray luminosity would not be driven by group mass, but by the thermodynamical state of the intragroup medium.

\begin{figure*}
    \centering
    {\includegraphics[width=0.49\textwidth]{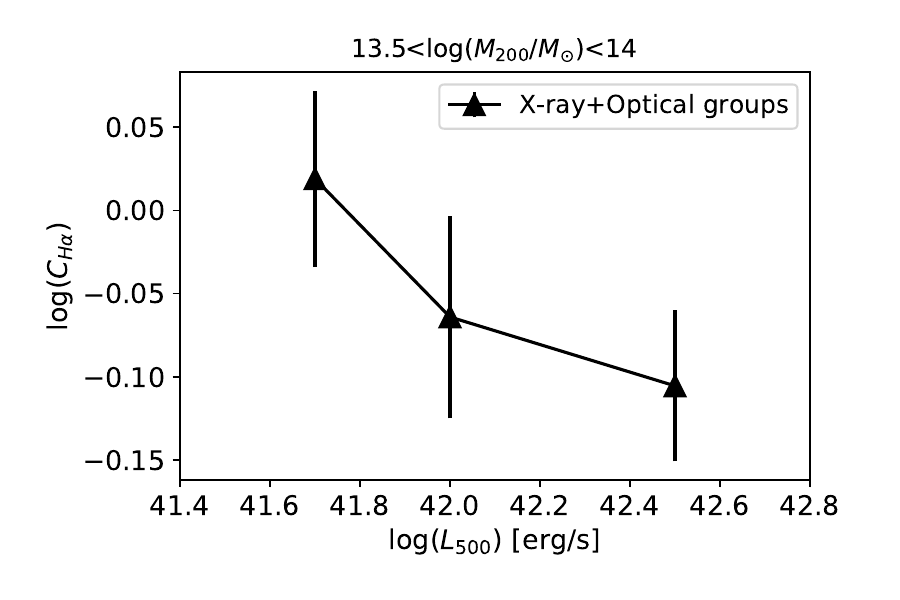}} 
    {\includegraphics[width=0.49\textwidth]{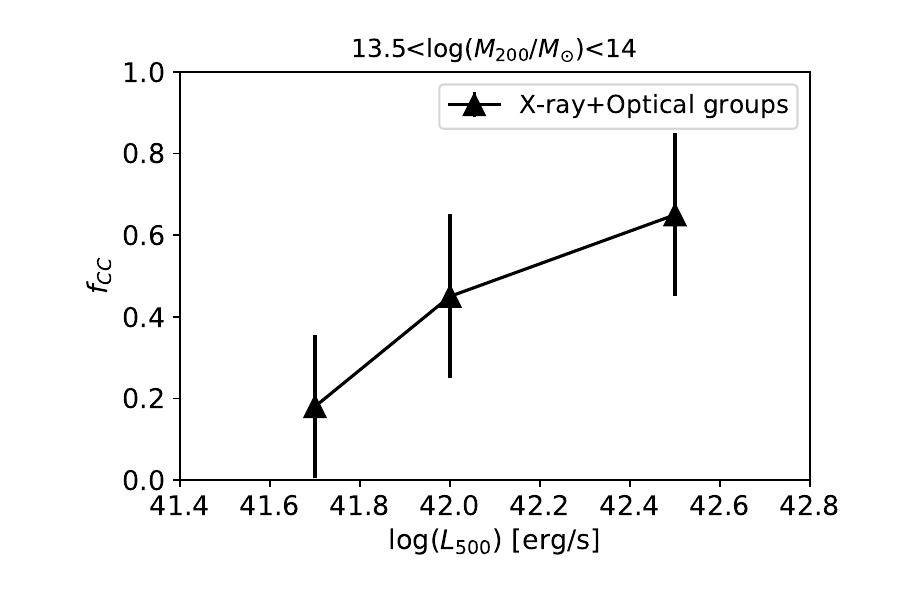}}
    \caption{Mean C-index (left) and fraction of centrally-concentrated star-forming galaxies (right) as a function of X-ray luminosity for 15 star-forming galaxies in 10 X-ray+Optical groups with $13.5<\log(M_{200}/M_{\odot})<14.5$. Star-forming galaxies in more X-ray luminous groups with the same dynamical mass show more concentrated star formation activity.}
    \label{Cindex_L500}
\end{figure*}

Figure~\ref{Cindex_L500} shows that the mean C-index decreases with increasing X-ray luminosity (left) and, consistently, the fraction of centrally-concentrated star-forming galaxies increases for higher X-ray luminosity (right). The Spearman rank correlation test (performed with 10,000 bootstrap samples to overcome the uncertainty due to the small sample size) implies a moderately significant anti-correlation between C-index and $\rm L_{500}$, with coefficient $\rho=-0.57$ and $p_{\rm S}=0.037$ (see Table~\ref{SpearmanResults}). Within X-ray+Optical groups, star-forming galaxies in more X-ray luminous groups with the same dynamical mass show more concentrated star formation activity. 

We note that X-ray+Optical groups are exclusively located within 2 Mpc of cosmic web nodes (Figure~\ref{LocalEnvironments_CW}), making it challenging to fully disentangle the effects of group thermodynamical state from those of node proximity. To assess whether the observed C-index–$L_{500}$ anti-correlation is driven primarily by proximity to nodes, we perform a partial Spearman rank correlation analysis controlling for $D_{\rm node}$. Once the dependence on node distance is accounted for, the correlation between C-index and $L_{500}$ is reduced but remains marginally significant (|$\rho|=0.55$, $p=0.046$). This reflects the fact that, within this restricted subsample of 10 X-ray+Optical groups, $L_{500}$ and $D_{\rm node}$ are only weakly correlated. These results suggest that the thermodynamical state of the intragroup medium may contribute to the observed trend in addition to the dominant role of node proximity, although disentangling node-driven effects from IGM-driven mechanisms conclusively will require larger samples.

\subsection{Spin--filament alignments for star-forming galaxies}
\label{Cindex and spin flips}

To constrain the physical processes regulating star formation concentration, we study stellar spin--filament alignments for regular star-forming galaxies and centrally-concentrated star-forming galaxies separately. We divide the 375 SAMI star-forming galaxies with measured spin--filament alignments into those with C-index $<-0.2$ (184) and those with C-index $>-0.2$ (191). Both sub-samples have galaxies spanning the range $9.5<\log(M_{\star}/M_{\odot})<12$.

\begin{figure*}
    \centering
    {\includegraphics[width=0.49\textwidth]{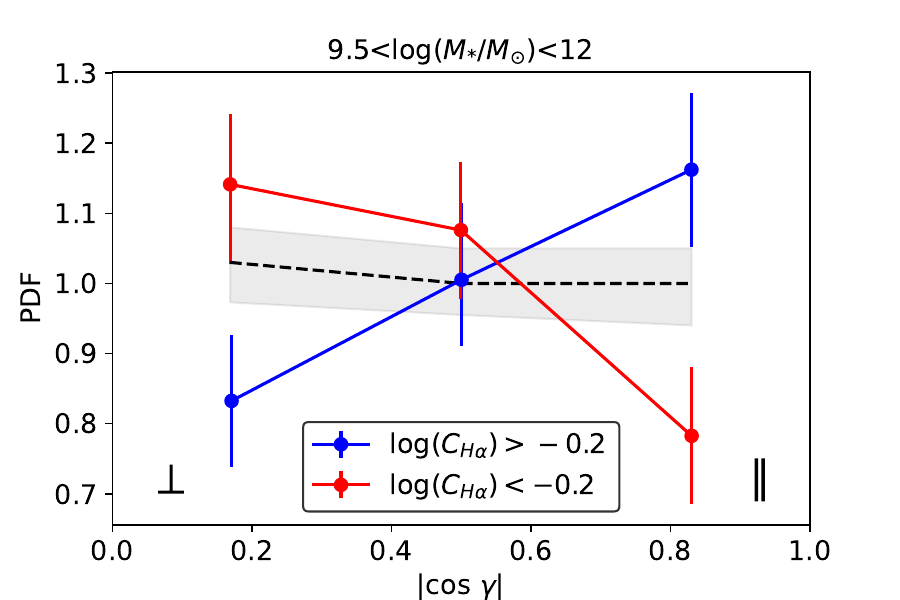}}
    {\includegraphics[width=0.49\textwidth]{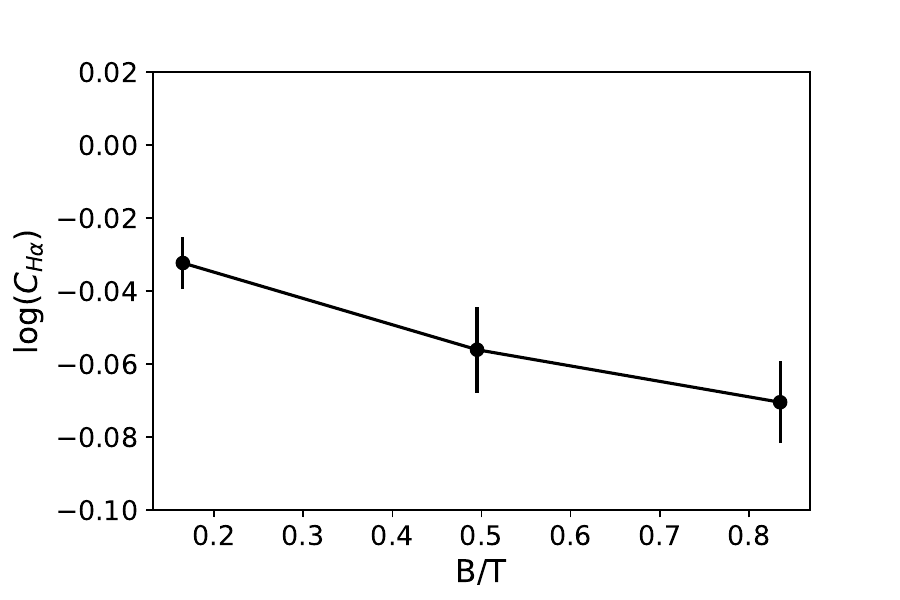}}
    \caption{Left panel: PDF of galaxy spin--filament alignments for 191 regular star-forming galaxies and 184 centrally-concentrated star-forming galaxies, with errors from 1000 bootstrap samples. The black dotted line and shaded region represent the null hypothesis and its 95\% confidence interval. The PDF is skewed towards more parallel spin--filament alignments (higher |$\cos\gamma$|) for regular star-forming galaxies (blue) and more perpendicular spin--filament alignments (lower |$\cos\gamma$|) for centrally-concentrated star-forming galaxies (red). Right panel: mean C-index as a function of bulge-to-total flux ratio. Bulge-dominated galaxies tend to have more centrally-concentrated star formation.}
    \label{Cindex_SpinFlips}
\end{figure*}

The left panel of Figure~\ref{Cindex_SpinFlips} shows the probability distribution function (PDF) in three bins of |$\cos\gamma$|, the absolute value of the cosine of the angle between the galaxy spin axis and the orientation of the closest filament. The PDFs are normalised so the mean over the bins is unity. The error bars are estimated from 1000 bootstrap samples. To assess the statistical significance of each trend, we apply the K-S test to the null hypothesis that |$\cos\gamma$| has a uniform distribution. To account for possible observational bias, we construct the null hypothesis by generating 3000 randomised galaxy samples with fixed spins but shuffled positions \citep[randomising the closest filaments;][]{Tempel2013a,Tempel2013b,Kraljic2021}. The median of the random samples is used as the null hypothesis; its PDF is nearly uniform. 

For regular star-forming galaxies we find that the PDF is skewed towards more parallel spin--filament alignments (higher |$\cos\gamma$|; $p_{\rm KS}=0.002$) and for centrally-concentrated star-forming galaxies towards more perpendicular spin--filament alignments (lower |$\cos\gamma$|; $p_{\rm KS}=0.005$). The |$\cos\gamma$| distribution of galaxies with C-index $<-0.2$ is also significantly different from that of galaxies with C-index $>-0.2$ ($p_{\rm 2KS}=0.001$). The stellar mass-matched analysis for the stellar spin-filament alignments reported in Appendix~\ref{Stellar mass-matched analysis of environmental trends} reveals the same conclusions: centrally-concentrated star-forming galaxies show more perpendicular alignments compared to mass-matched regular star-forming galaxies. Finally, consistent results are also found for the ionised gas--spin filament alignments for 635 SAMI galaxies in the range $7.5<\log(M_{\star}/M_{\odot})<12$. 

Since galaxy spin--filament alignments are found to be linked to bulge growth \citep{Barsanti2022}, we explore (in the right panel of Figure~\ref{Cindex_SpinFlips}) C-index as a function of bulge-to-total flux ratio. As B/T increases, C-index significantly decreases ($\rho=-0.18$, $p_s=10^{-3}$), meaning that more bulge-dominated galaxies tend to have more centrally-concentrated star formation.

These results suggest regular star-forming galaxies maintain a parallel spin--filament alignment that sustains extended star formation activity via coherent gas accretion from filaments. They also tend to be disc-dominated galaxies. By contrast, the more perpendicular spin--filament alignment for centrally-concentrated star-forming galaxies suggests these galaxies have likely undergone significant dynamical processing (mergers, environmental interactions, starbursts or gas accretion) that reorients spin and centrally concentrates star formation. As galaxies evolve toward being more bulge-dominated, their star formation becomes more centrally-concentrated.

\section{Discussion}
\label{Discussion}

Here we discuss the possible physical interpretations of our results and compare our findings to previous work. Finally, we address the limitations due to the small galaxy samples available in this study and the interpretation of the C-index.

\subsection{Ram-pressure stripping in X-ray+Optical groups}
\label{RPS in X-ray+Optical groups}

In Section~\ref{Concentrated star formation in local environments} we found that centrally-concentrated star formation and outside-in quenching is more common in X-ray+Optical groups than in Optical groups and the Field, independent of stellar mass or group mass (lowest median C-index and highest $f_{\rm CC}$). The consistent trends across stellar mass bins suggests environmental effects, rather than internal processes regulated by stellar mass, are the dominant driver of the observed central concentration of star formation in X-ray+Optical groups. These results agree with the conclusions of \citet{Schaefer2019} and \citet{Wang2022}, who studied the C-index for SAMI star-forming galaxies in local environments using different (but related) local metrics.

Within X-ray+Optical groups, star-forming galaxies in more X-ray luminous systems with the same dynamical mass exhibit more centrally-concentrated star formation activity (Figure~\ref{Cindex_L500}). Since X-ray luminosity traces the density of the IGM, this suggests a denser IGM more effectively removes gas from the outskirts of galaxies, confining star formation to the central regions. This may be interpreted as ram-pressure stripping acting in X-ray+Optical groups to drive outside-in quenching. The idea that ram-pressure stripping, which acts on short timescales of $\sim$0.5--1\,Gyr to remove the outer disc gas and stop star formation in the galaxy outskirts, is the dominant environmental mechanism in X-ray+Optical groups is supported by other studies. \citet{Wang2022} analyse centrally-concentrated star-forming galaxies as a function of stellar age, D$_{\rm n}$4000 index, and H$\delta_{\rm A}$ index, concluding that outside-in quenching is effective in groups but the process does not rapidly quench the whole galaxy. For a deeper understanding of the star formation quenching timescales, Wang et~al.\ (in prep.) investigate star formation histories of centrally-concentrated star-forming galaxies in different environments. They conclude that the quenching timescale is shorter for high-mass groups ($12.5<\log(M_{200}/M_{\odot})<14$) than for low-mass groups ($\log(M_{200}/M_{\odot})\leq12.5$) and ungrouped regions, highlighting that  quenching is faster in denser regions. Since X-ray+Optical groups are the most massive and dense environments of those studied, we expect them to exhibit relatively rapid quenching, consistent with ram-pressure stripping.

We note that while the observed trends in X-ray+Optical groups are consistent with environmental processes such as ram-pressure stripping, they do not uniquely demonstrate this mechanism. The specific physical driver cannot be unambiguously identified from the current observations. Other processes, including tidal interactions, gas inflows, or centrally triggered starbursts, could contribute to centrally-concentrated star formation. We address the caveats regarding the interpretation of C-index as an outside-in quenching tracer in Section~\ref{Interpretations of C-index}.

Finally, there is a potential degeneracy between local and large-scale environmental effects. X-ray+Optical groups in our sample are preferentially located in close proximity to cosmic nodes (Figure~\ref{LocalEnvironments_CW}), and we find that the distance to the nearest node is the dominant environmental parameter regulating the C-index (Section~\ref{Nodes shape star formation concentration}). As a result, the enhanced central concentration of star formation observed in X-ray+Optical groups may partially reflect their location within dense node environments, rather than the presence of hot intragroup gas alone. While our partial correlation analysis in Section~\ref{Correlation of C-index with X-ray luminosity} suggests that the thermodynamical state of the intragroup medium may contribute to the observed trend in addition to the dominant role
of node proximity, the strong spatial correlation between X-ray+Optical groups and nodes in the current sample prevents a clean separation of these influences. We therefore interpret the observed trends as arising from a combination of local (e.g. intragroup medium) and large-scale (node proximity) environmental processes, supporting a scenario where environment act on multiple scales to concentrate star formation toward the centres of galaxies.

\subsection{Nodes shape star formation concentration}
\label{Nodes shape star formation concentration}

X-ray+Optical galaxy groups are found closer to nodes (Figure~\ref{LocalEnvironments_CW}), in agreement with \citet{Popesso2024}, who investigate the properties of the GAMA galaxy groups detected and undetected by eROSITA, using different techniques to cross-match the GAMA and eROSITA data and to reconstruct the cosmic web \citep{Eardley2015}. 

In line with X-ray+Optical galaxy groups showing strong centrally concentrated star formation, we find that star-forming galaxies located in Nodes exhibit the lowest median C-index compared to those in Filaments and Voids (Figure~\ref{Cindex_SF_CW}). Two-sample K-S tests confirm that C-index distributions differ significantly between Nodes and the other large-scale environments for galaxies with stellar masses $7.5<\log(M_\star/M_\odot)<12$, but not for the subsample of high-mass galaxies with $\log(M_\star/M_\odot) > 9.5$ (Table~\ref{Cindex_KS_CW}). This suggests that a more rapid or more efficient outside-in quenching occurs for low-mass galaxies ($7.5<\log(M_\star/M_\odot)<9.5$), which have shallower potential wells and so are more susceptible to external influences in dense node environments, whereas quenching in massive galaxies is more governed by internal processes and secular evolution. Such a conclusion is consistent with the study of \citet{Pan2025}, who analyse 3098 low-redshift star-forming galaxies from the MaNGA spatially-resolved galaxy survey \citep{Bundy2015}, finding that low-mass satellite galaxies below the main sequence tend to have higher concentration. We do not observe this stellar mass-dependent trend for star formation concentration within local environments, however, highlighting the distinct role that the cosmic web plays in shaping galaxy evolution.

The decrease in C-index with decreasing distance from nodes (and so increasing distance from voids), coupled with no trend with distance to filaments, suggests that the very dense (or under-dense) cosmic regions play a role in the central concentration of star formation in star-forming galaxies, while filaments play a more neutral or transitional role (Figure~\ref{Cindex_Env}). Within filaments, galaxies might have stable or replenished gas reservoirs, allowing more extended star formation to persist. In these intermediate-density environments, outside-in quenching might not be efficient enough to leave a signature in the galaxy. The lack of a trend in C-index with $D_{\rm fil}$ suggests that filaments neither strongly compress nor strip gas in a way that preferentially concentrates or quenches star formation centrally. 

No significant residual correlation is left between C-index and any other environment metrics when controlling for $D_{\rm node}$ (Figure~\ref{PartialCorrelations}). Thus, proximity to nodes is key to shaping star formation concentration in star-forming galaxies, pointing to the strongest gas-related environmental mechanisms causing outside-in quenching occurring close to nodes.

While for star formation concentration we find a dominant role for nodes with respect to filaments, \citet{Barsanti2025} find that filaments play a key role in slowing down stellar spin compared to nodes. The pre-processing by mergers occurring within filaments is a strong enough process to slow down stellar spin and start the formation of slow-rotating galaxies before they reach nodes; but it is the environmental mechanisms occurring close to and within nodes, such as ram-pressure stripping, that are able to affect the star formation distribution within the galaxy. Together, these results point to dual imprints of the cosmic web on nearby galaxies: filaments leave a kinematic signature, while nodes leave a mark on star formation. 

\subsection{Concentration of star formation and bulge growth}
\label{Concentration of star formation and bulge growth}

The spatial distribution of star formation in galaxies is closely linked to their internal structure, with increasing central star formation associated with bulge growth and the build-up of stellar mass in the inner regions \citep{Ellison2018,Pan2025}. The growth of the bulge component within the galaxy is also linked to the flipping of galaxy spin--filament alignments, shifting from a parallel to a more perpendicular configuration \citet{Barsanti2022}.

In agreement with these studies, we find that regular star-forming galaxies tend to have their spins aligned in parallel with the direction of the closest filament (Figure~\ref{Cindex_SpinFlips}), pointing to gas accretion from the material of the filament to build up angular momentum and sustain extended disc-like star formation. On the other hand, centrally-concentrated star-forming galaxies show a more perpendicular tendency for their spin--filament alignments, suggesting that strong interactions such as mergers, outside-in environmental quenching processes, or centralised starbursts have decoupled their angular momentum from the filament direction. These mechanisms funnel gas into the central galaxy regions, enhancing star formation in the core and driving the build-up of the bulge component. This interpretation is further supported by the strong correlation observed between C-index and bulge-to-total flux ratio, with bulge-dominated galaxies showing more centrally-concentrated star formation.

\subsection{Limitations due to small galaxy samples}
\label{Limitations due to small galaxy samples}

We note that, although we recover statistically significant results, this study is limited by small galaxy samples. The combination of the GAMA group catalogue with the eROSITA data and the SAMI Galaxy Survey allows us to build a powerful dataset for studying spatially-resolved star formation quenching in a variety of local environments and within the large-scale structure, but the galaxy and group samples we can construct are small. In particular, the number of X-ray+Optical groups drops from 79 to 20 with available SAMI data (see Section~\ref{Characterisation of local environments}) and, due to the many criteria applied for reliable C-index measurements (see Section~\ref{Characterisation of star-forming galaxies}), we recover only 22 star-forming galaxies in the 20 X-ray+Optical groups. This limited number of galaxies in X-ray+Optical groups introduces significant statistical uncertainty. The observed trends should therefore be interpreted as suggestive rather than conclusive, and the future larger samples outlined below will be required to robustly confirm these results.

Moreover, G09 is the GAMA region with the most (13 out of 20) X-ray+Optical groups relative to the G12 and G15 regions. This is because the eROSITA/eFEDS data have a depth equivalent to about ten all-sky survey passes in the G09 region, but only one all-sky pass in the G12 and G15 regions (Section~\ref{eROSITA eFEDS and eRASS1 data}), and highlights the impact that the depth of the X-ray data has on characterising environments. It is also important to note that X-ray selection itself introduces systematic effects: the source detection pipeline fits an extended $\beta$-surface brightness model and therefore preferentially identifies morphologically concentrated intracluster media, while very extended or irregular groups may be missed. As a result, the X-ray+Optical group sample used here is relatively low in completeness but highly pure.

The vast upcoming 4MOST spectroscopic redshift surveys for millions of galaxies, such as the Wide Area VISTA Extra-galactic Survey (WAVES; \citealp{Driver2016,Kaur2025}) and the 4MOST Hemisphere Survey of the nearby Universe (4HS; \citealp{Taylor2023}), will be extremely complete. They can be combined with next-generation IFS galaxy surveys, such as the Hector survey \citep{Bryant2020,Bryant2024,Oh2025} that will observe 15,000 galaxies out to 2 effective radii, and with future releases of eROSITA data, having greater depth and more all-sky passes, to build larger and more powerful galaxy and group samples to study spatially-resolved star formation quenching at different environmental scales.

\subsection{Interpretations of C-index}
\label{Interpretations of C-index}

In this work we interpret C-index as a tracer of outside-in quenching, motivated by the results of \citet{Wang2022, Wang2023}. Using SAMI star-forming galaxies, which include our sample, \citet{Wang2022} analyse radial profiles of light-weighted age, mass-weighted age, D$_{\rm n}4000$, and H$\delta_{\rm A}$. They find that centrally-concentrated star-forming galaxies in high-mass groups exhibit younger inner regions and older outer discs compared to ungrouped galaxies, consistent with enhanced central star formation and earlier quenching in the outskirts. In contrast, galaxies in lower-mass groups or ungrouped environments show shallower age gradients, suggesting that centrally concentrated star formation can also arise from secular processes.

Additional support comes from \citet{Wang2023}, who study environmental quenching in the EAGLE/C-EAGLE simulations over the last 11\,Gyr \citep{Schaye2014,Barnes2017} and compare to SAMI observations at $z=0$. Galaxies with low C-index values tend to lie below the star formation rate--stellar mass main sequence and exhibit declining specific star formation rate profiles with increasing radius, consistent with outside-in quenching. Together, these results support the interpretation that the C-index can be used as a tracer of outside-in quenching for the SAMI star-forming sample.

However, the C-index alone does not uniquely diagnose the physical mechanism driving centrally concentrated star formation. Centrally enhanced H$\alpha$ emission may also arise from processes such as mergers, bar-driven inflows or tidal interactions, which can increase central star formation without necessarily suppressing the outer disc as in the case of ram-pressure stripping. We therefore emphasise that while the observed trends in Sections~\ref{Concentrated star formation in local environments} and \ref{Correlation of C-index with X-ray luminosity} are consistent with outside-in quenching and ram-pressure stripping, they do not demonstrate that this is the sole driver.

\section{Summary and conclusions}
\label{Summary and conclusions}

In this work, we explore star formation concentration (C-index$=\log{(r_{50,{\rm H}\alpha}/r_{50,\rm cont})}$) for star-forming galaxies within different local and large-scale environments combining the three datasets: (i)~the GAMA spectroscopic redshift survey for optical selection of galaxy groups and reconstruction of the cosmic web; (ii)~the eROSITA data within the GAMA regions to identify optically-selected groups with X-ray detection; and (iii)~the SAMI Galaxy Survey to characterise spatially-resolved star formation. 

We categorise local environments into X-ray+Optical groups, Optical groups (no X-ray counterpart), and the Field. We use {\sc DisPerSE} to trace the cosmic web, measuring for each SAMI galaxy the distances to the closest Node, Filament and Void, and thus associate it with one of these large-scale structures. Our SAMI sample consists of 649 star-forming galaxies with $7.5<\log(M_{\star}/M_{\odot})<12$, with 22 in X-ray+Optical groups, 276 in Optical groups, and 351 in the Field. With respect to the large-scale structure, 126 star-forming galaxies belong to Nodes, 211 to Filaments and 312 to Voids. Our main results are:
\begin{enumerate}[left=-3pt .. \parindent]
\item Star-forming galaxies in X-ray+Optical groups show the lowest median C-index and the highest percentage of star-forming galaxies with centrally concentrated star formation relative to Optical groups and the Field, independently of group mass or stellar mass.  
\item Star-forming galaxies in Nodes show the lowest median C-index and the highest percentage of centrally-concentrated star-forming galaxies relative to Filaments and Voids, driven by low-mass galaxies. Filaments and Voids do have significantly different C-index distributions.
\item C-index correlates most significantly with distance to the closest Node ($D_{\rm node}$), with no role left for other local or large-scale environment metrics ($\Sigma_5$, $D_{\rm fil}$, $D_{\rm void}$).
\item X-ray+Optical groups are found at Nodes. Within X-ray+Optical groups with the same dynamical mass, star-forming galaxies in more X-ray luminous groups show more concentrated star formation activity.
\item Regular star-forming galaxies tend to have spins aligned parallel to filaments, while centrally-concentrated star-forming galaxies tend to have spins aligned perpendicular to filaments.
\end{enumerate}  

\noindent In conclusion, our results support a scenario where environmental processes at multiple scales act to concentrate star formation toward the centres of galaxies, driving bulge growth and altering galaxy structure. The most centrally concentrated star-forming galaxies are found in the densest environments, X-ray+Optical groups and Nodes, suggesting that interactions and processes unique to these regions (such as ram-pressure stripping, tidal interactions, strangulation, or galaxy mergers) remove or heat the gas in the outskirts while leaving the central gas reservoirs intact or compressed. This leads to a spatially contracted distribution of star formation. The correlation between C-index and $D_{\rm node}$ further supports the idea that proximity to gravitational potential minima, where dynamical interactions and gas processing are enhanced, is a key driver of central star formation.

The link between spin--filament alignments and star formation concentration offers insight into how the cosmic web influences internal galaxy structure. Galaxies with centrally-concentrated star formation show a tendency toward perpendicular spin--filament alignments, potentially reflecting a history of angular momentum reorientation through mergers contributing to the build-up of the bulge. This suggests that the cosmic web not only shapes the spatial distribution of galaxies but also affects their internal kinematics and star formation geometry. 

We note that a centrally concentrated $\rm H\alpha$ distribution, reflected by low C-index values, is not produced uniquely by a single physical mechanism. In addition to environmentally driven outside-in quenching (e.g. ram-pressure stripping), centrally concentrated star formation may also result from central starbursts triggered by gas inflows, major or minor mergers, or residual AGN/LINER-like emission. Although we applied the \citet{Wang2022} correction to mitigate non–star-forming ionization, additional diagnostics are not available for the full sample and some level of ambiguity remains. This caveat should be kept in mind when interpreting C-index measurements.

Future studies will be essential to gain a more complete understanding of the physical processes shaping star formation concentration. In particular, building larger and more diverse galaxy samples, such as those enabled by 4MOST and Hector, will allow for a clearer disentanglement of the relative roles of environmental mechanisms. Surveys targeting different redshift ranges, like MAGPI \citep{Foster2021}, and those designed to probe unbiased regions of the sky with respect to large-scale environment, such as MAGNET (Vulcani et al., in preparation), will be crucial to tracing the evolution of environmental processes across cosmic time. The combination of these data sets, alongside upcoming eROSITA data releases, will provide a powerful framework for addressing these open questions.

\section*{Acknowledgements}
We thank the referee for the constructive reports. This research was supported by the Australian Research Council Centre of Excellence for All Sky Astrophysics in 3 Dimensions (ASTRO~3D, CE170100013). The SAMI Galaxy Survey is based on observations made at the Anglo-Australian Telescope. The Sydney-AAO Multi-object Integral field spectrograph (SAMI) was developed jointly by the University of Sydney and the Australian Astronomical Observatory, and funded by ARC grants FF0776384 (Bland-Hawthorn) and LE130100198. The SAMI input catalogue is based on data taken from the Sloan Digital Sky Survey, the GAMA Survey, and the VST/ATLAS Survey. The SAMI Galaxy Survey website is http://sami-survey.org/. 

This work is based on data from eROSITA, the soft X-ray instrument aboard
SRG, a joint Russian-German science mission supported by the Russian Space
Agency (Roskosmos), in the interests of the Russian Academy of Sciences represented
by its Space Research Institute (IKI), and the Deutsches Zentrum f\"ur Luft
und Raumfahrt (DLR). The SRG spacecraft was built by Lavochkin Association
(NPOL) and its subcontractors and is operated by NPOL with support from the
Max Planck Institute for Extraterrestrial Physics (MPE).
The development and construction of the eROSITA X-ray instrument were led
by MPE, with contributions from the Dr. Karl Remeis Observatory Bamberg
\& ECAP (FAU Erlangen-Nuernberg), the University of Hamburg Observatory,
the Leibniz Institute for Astrophysics Potsdam (AIP), and the Institute for Astronomy and Astrophysics of the University of T\"ubingen, with the support of
DLR and the Max Planck Society. The Argelander Institute for Astronomy of the University of Bonn and the Ludwig Maximilians Universität Munich also
participated in the science preparation for eROSITA. The eROSITA data used here were processed using the eSASS/NRTA software system developed by the German eROSITA consortium.

SB acknowledges the support from the Physics Foundation through the Messel Research Fellowship. AL acknowledges the support from the National Natural Science Foundation of China (Grant No. 12588202). AL is supported by the China Manned Space Program with grant no. CMS-CSST-2025-A04. EB and AL acknowledge financial support from the European Research Council (ERC) Consolidator Grant under the European Union’s Horizon 2020 research and innovation program (grant agreement CoG DarkQuest No 101002585). SO acknowledges support from the Korean National Research Foundation (No. RS-2023-00214057 and No. RS-2025-00514475). JJB acknowledges support of an Australian Research Council Future Fellowship (FT180100231). AR recognises the support from the Australian Research Council Centre of Excellence in Optical Microcombs for Breakthrough Science (project number CE230100006), funded by the Australian Government.

This study uses data provided by AAO Data Central (http://datacentral.org.au/) and the \href{http://www.python.org}{Python} programming language \citep{vanrossum1995}. We acknowledge the use of {\sc \href{https://pypi.org/project/numpy/}{numpy}} \citep{harris+2020}, {\sc \href{https://pypi.org/project/scipy/}{scipy}} \citep{jones+2001}, {\sc \href{https://pypi.org/project/matplotlib/}{matplotlib}} \citep{hunter2007}, {\sc \href{https://pypi.org/project/astropy/}{astropy}} \citep{astropyco+2013}, {\sc \href{https://pingouin-stats.org/}{pingouin}} \citep{Vallat2018}, {\sc \href{https://scikit-learn.org/stable/}{scikit-learn}} \citep{Pedregosa2012} and {\sc \href{http://www.star.bris.ac.uk/~mbt/topcat/}{topcat}} \citep{taylor2005}.

\section*{Data availability}

The SAMI data used in this paper are publicly available at \href{https://docs.datacentral.org.au/sami}{SAMI Data Release 3} \citep{Croom2021}. Ancillary data are from \href{http://gama-survey.org}{GAMA Data Releases 3 \& 4} \citep{Baldry2018,Driver2022}. The X-ray group and cluster catalogue from eROSITA/eFEDS is available at \href{https://erosita.mpe.mpg.de/edr/eROSITAObservations/Catalogues/}{eROSITA-DE: Early Data Release site} \citep{Liu2022} and the eROSITA/eRASS1 group and cluster catalogue is available at \href{https://erosita.mpe.mpg.de/dr1/AllSkySurveyData_dr1/Catalogues_dr1/}{eROSITA-DE: Data Release 1 site} \citep{Bulbul2024}.

\section*{Author Contribution Statement}

SB devised the project, carried out the analysis and drafted the paper. DW estimated C-index measurements. AL and EB provided preliminary eROSITA catalogues. MO provided spatially-resolved spectroscopic classification. SB, MO, MC, SMC and BV contributed to data analyses and interpretation of the results. All authors discussed the results and commented on the manuscript.  


\bibliographystyle{mnras}
\bibliography{biblioSAMI} 


\appendix

\section{Sensitivity of the optical and X-ray group matching}
\label{Sensitivity of the optical and Xray group matching}

To test the sensitivity of our results to the adopted matching radius between optical GAMA groups and eROSITA-detected groups, we repeat the cross-matching using radii of $60^{\prime\prime}$ and $180^{\prime\prime}$, bracketing our fiducial choice of $120^{\prime\prime}$. A radius of $60^{\prime\prime}$ is comparable to the on-axis eROSITA PSF and provides a conservative lower limit, while $180^{\prime\prime}$ allows for larger centroid offsets due to off-axis PSF broadening, substructure, or asymmetric X-ray emission.

\begin{table*}
\caption{Sensitivity of the optical and X-ray group matching to the adopted angular radius. Column~1 lists the matching radius, column~2 the number of X-ray+Optical groups, columns~3 the number of X-ray+Optical groups with SAMI galaxies, and column~4 the number of SAMI galaxies within those groups.}
\label{MatchingRadius}
\centering
\begin{tabular}{lccc}
\toprule
Matching radius & X-ray+Optical groups & X-ray+Optical groups with SAMI & SAMI galaxies \\
\midrule
$60^{\prime\prime}$  & 61 & 17 & 70 \\
$120^{\prime\prime}$ (fiducial) & 79 & 20 & 77 \\
$180^{\prime\prime}$ & 93 & 22 & 80 \\
\bottomrule
\end{tabular}
\end{table*}

Table~\ref{MatchingRadius} summarises the number of optical groups matched to eROSITA detections for each choice. While the total number of matched groups increases with matching radius, the subset of X-ray+Optical groups hosting SAMI galaxies and the number of these latter are largely stable. Repeating the analysis with these alternative matching radii yields consistent median C-index values and centrally-concentrated star-forming fractions within the quoted uncertainties. We therefore conclude that our main results are robust against reasonable variations in the matching radius.

\section{Impact of flux-limited selection and redshift-dependent sampling}
\label{Impact of flux-limited selection and redshift-dependent sampling}

\begin{figure}
    \centering
    {\includegraphics[width=\columnwidth]{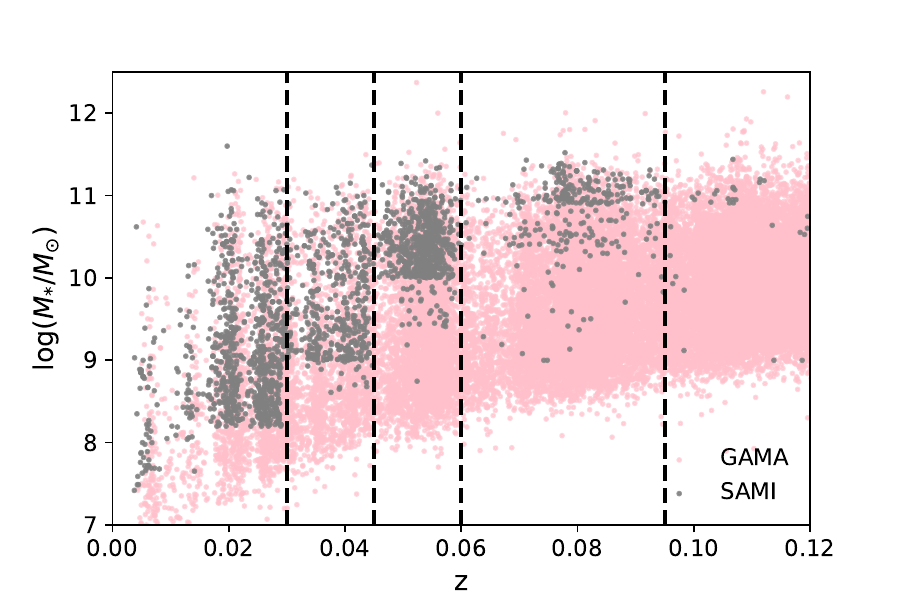}}
    \caption{Stellar mass vs. redshift distribution for GAMA galaxies (pink) and SAMI galaxies (grey). The dashed black lines mark the SAMI selection tiers at $z = 0.03, 0.045, 0.06, 0.095$.}
    \label{GAMA_z_vs_Mstar}
\end{figure}

The GAMA and SAMI galaxy catalogues are not strictly volume-limited, and their effective number density decreases with redshift. The SAMI Galaxy Survey employs stepped redshift and stellar mass selection tiers at $z = 0.03, 0.045, 0.06, 0.095$, which introduce an intrinsic mass–redshift coupling by design \citep{Bryant2015}. Figure~\ref{GAMA_z_vs_Mstar} shows the stellar mass vs. redshift distribution for GAMA and SAMI galaxies, illustrating that the GAMA tracer population becomes increasingly dominated by higher-mass systems with redshift, while low-mass SAMI galaxies with $\log(M_{\star}/M_{\odot})<10$ are largely absent at the higher-redshift tiers.

To mitigate the impact of this inherent selection bias on the cosmic web analysis, we perform two robustness tests. First, we verify that our main results are unchanged when restricting the cosmic web reconstruction to a stellar mass–limited GAMA subsample that is approximately complete for $\log(M_{\star}/M_{\odot})\geq9.5$ over $0<z\leq0.12$. Second, we repeat the C-index analysis in two redshift intervals aligned with the SAMI selection tiers ($0<z\leq0.045$ and $0.045<z\leq0.12$) without imposing any additional stellar mass cut. This tier-based robustness test closely follows the approach adopted by \citet{Welker2020} in their cosmic web analysis of GAMA galaxies combined with SAMI data, where the analysis was repeated independently within each SAMI redshift tier, finding stable large-scale results.

In both tests, the large-scale environmental trends in C-index remain consistent within the uncertainties with those presented in Section~\ref{Concentrated star formation in the cosmic web}: Nodes have the lowest median C-index value and the highest fraction of centrally-concentrated star-forming galaxies, while no significant statistical difference is measured for the C-index distributions of Filaments and Voids. We additionally confirm that the C-index–$D_{\rm node}$ correlation remains statistically significant when controlling for redshift via a partial Spearman test ($|\rho|=0.1$, $p=0.009$), indicating that our conclusions are not driven by redshift-dependent sampling.

\section{Sensitivity of the cosmic web reconstruction to the persistence threshold}
\label{Sensitivity of the Cosmic Web Reconstruction to the Persistence Threshold}

We adopt a 3$\sigma$ persistence threshold in the cosmic web reconstruction from {\sc DisPerSE}. This choice efficiently suppresses low-persistence features most sensitive to sampling noise and survey boundaries, while retaining the dominant, physically meaningful filamentary structures traced by the galaxy distribution and it is widely used for observational cosmic web analysis \citep[e.g.][]{Sousbie2011a,Welker2020,Kraljic2021,Barsanti2022}.

To assess the robustness of the cosmic web reconstruction to the choice of persistence threshold, we repeat the {\sc DisPerSE} analysis using thresholds of 2$\sigma$, 3$\sigma$ (fiducial), and 4$\sigma$. As expected, lowering the threshold increases the number of detected filaments and nodes by including lower-persistence features, while raising it retains only the most prominent structures.

Despite these differences in the absolute number of filament segments ({\sc DisPerSE} arcs) and nodes (reported in Table~\ref{tab:persistence_test}), the association of the SAMI star-forming galaxies within the large-scale environments and the resulting C-index distributions remain stable across the three thresholds. Only $\leq5$\% of galaxies change large-scale environment classification when varying the persistence threshold, with most re-classifications occurring between Filaments and Voids; Node membership remains the most robust. The relative ordering of C-index medians between Nodes, Filaments, and Voids, as well as the trends with node distance, are unchanged within the uncertainties. This demonstrates that our conclusions are not sensitive to the specific choice of persistence threshold.

\begin{table}
\caption{The number of detected filament segments and nodes is shown for persistence thresholds of 2$\sigma$, 3$\sigma$ (fiducial), and 4$\sigma$.}
\centering
\begin{tabular}{lcc}
\toprule
Persistence threshold & N$_{\rm filament\,segments}$ & N$_{\rm nodes}$ \\
\midrule
2$\sigma$ & 1088 & 527 \\
3$\sigma$ & 957  & 489 \\
4$\sigma$ & 866  & 462 \\
\bottomrule
\end{tabular}
\label{tab:persistence_test}
\end{table}

\section{Stellar mass-matched analysis of environmental trends}
\label{Stellar mass-matched analysis of environmental trends}

To remove any residual dependence on stellar mass, we perform a mass-matching analysis when comparing different environments. For each comparison, we construct subsamples with identical stellar mass distributions by randomly resampling the larger population to match the mass distribution of the smaller one. This procedure is repeated 1000 times to account for sampling noise.

\begin{figure}
    \centering
    {\includegraphics[width=\columnwidth]{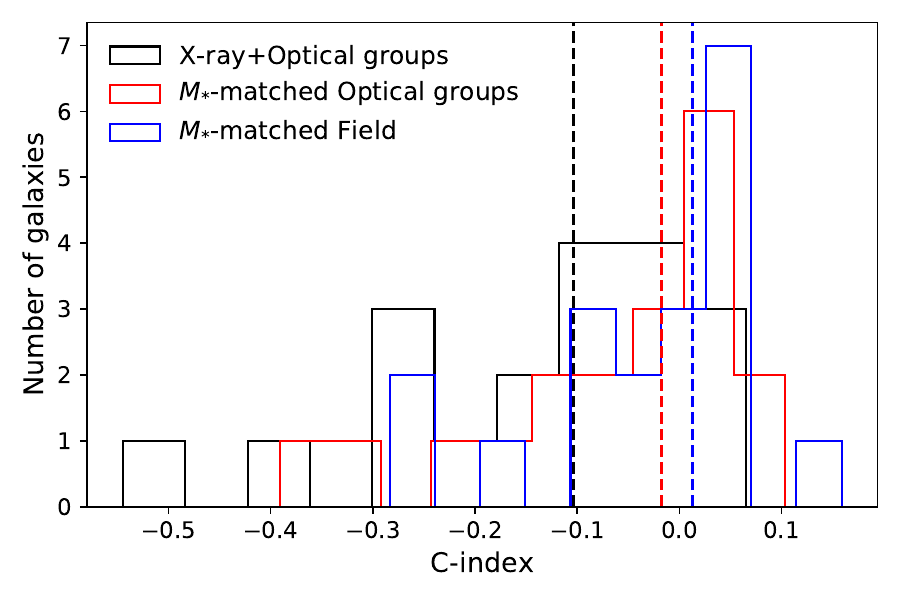}}
    {\includegraphics[width=\columnwidth]{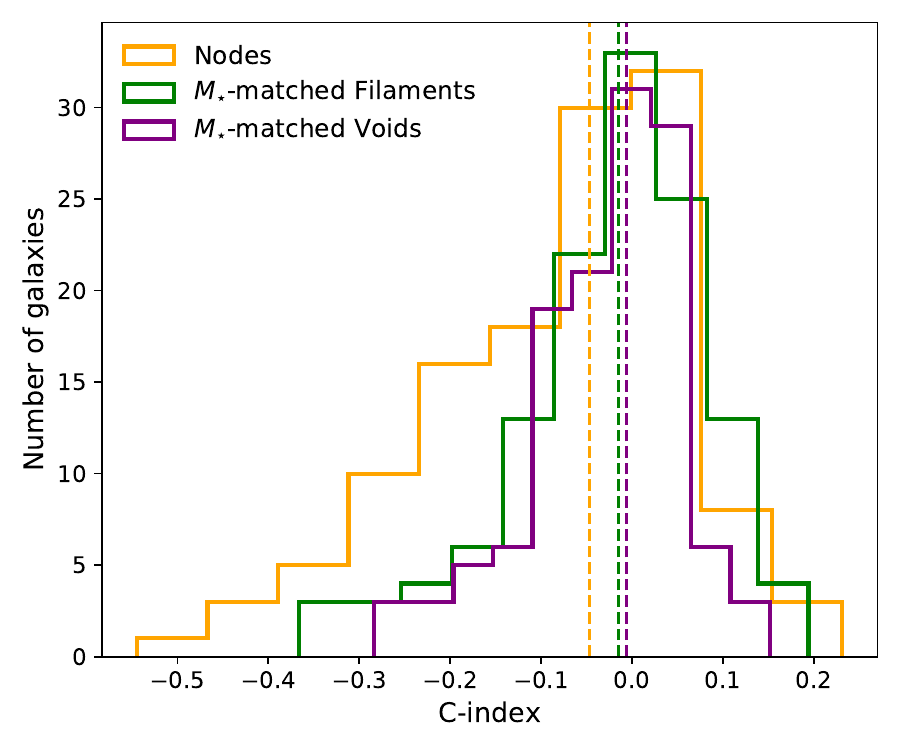}}
    \caption{Top panel: C-index distributions for star-forming galaxies in X-ray+Optical groups (black), stellar mass-matched Optical groups (red), and stellar mass-matched Field (blue). The median C-index (dashed line) is still lowest for X-ray+Optical groups. Bottom panel: C-index distributions for star-forming galaxies in Nodes (orange), stellar mass-matched Filaments (green), and stellar mass-matched Voids (purple). The median C-index (dashed line) is still lowest for Nodes.}
    \label{Mass_matched_Cindex}
\end{figure}

After matching the stellar-mass distributions, the C-index distributions in X-ray+Optical groups remain systematically shifted toward lower values compared to Optical groups and the Field, as shown in the top panel of Figure~\ref{Mass_matched_Cindex}. In the mass-matched X-ray+Optical versus Optical group comparison, the effect size is stable with a median Cliff’s delta of $\delta=-0.35$ with a 95\% C.I. of $(-0.47,-0.23)$, while the KS test becomes less significant with median 
$p_{\rm 2KS}=0.15$ and a 16th–84th percentile range of (0.03,0.54), due to the reduced sample size (22 galaxies in each sample). In contrast, the mass-matched X-ray+Optical groups versus Field comparison remains statistically significant, with a stronger effect size 
$\delta=-0.51$ with a 95\% C.I. of $(-0.61,-0.41)$ and median 
$p_{\rm 2KS}=0.027$ with a 16th–84th percentile range of (0.003,0.068). This confirms that the enhanced central concentration of star formation in X-ray+Optical groups is not driven by stellar mass differences but reflects a genuine environmental effect.

Similarly, the lower C-index values observed in Nodes compared to Filaments and Voids persist in the mass-matched samples shown in the bottom panel of Figure~\ref{Mass_matched_Cindex}, indicating that the large-scale environmental trends are not a by-product of stellar mass segregation. When Filaments are resampled to match the mass distribution of Nodes, we find a median KS p-value of 
$p_{\rm 2KS}=0.0039\,(0.0009,0.014)$ and a stable effect size of 
$\delta=-0.23\,(-0.26,-0.20)$, confirming a systematic shift toward lower C-index values in Nodes. The Nodes versus Voids comparison remains significant after mass matching, with $p_{\rm 2KS}=0.0060\,(0.0015,0.021)$ and $\delta=-0.20\,(-0.23,-0.17)$, indicating that galaxies in Nodes have more centrally concentrated star formation than those in Voids even at fixed stellar mass. The Filaments versus Voids comparison shows no significant difference after mass matching, with a median 
$p_{\rm 2KS}=0.43\,(0.30,0.50)$ and a negligible effect size 
$\delta=+0.02\,(0.00,0.04)$, in agreement with the results for the unmatched samples in Section~\ref{Concentrated star formation in the cosmic web}.

\begin{figure}
    \centering
    {\includegraphics[width=\columnwidth]{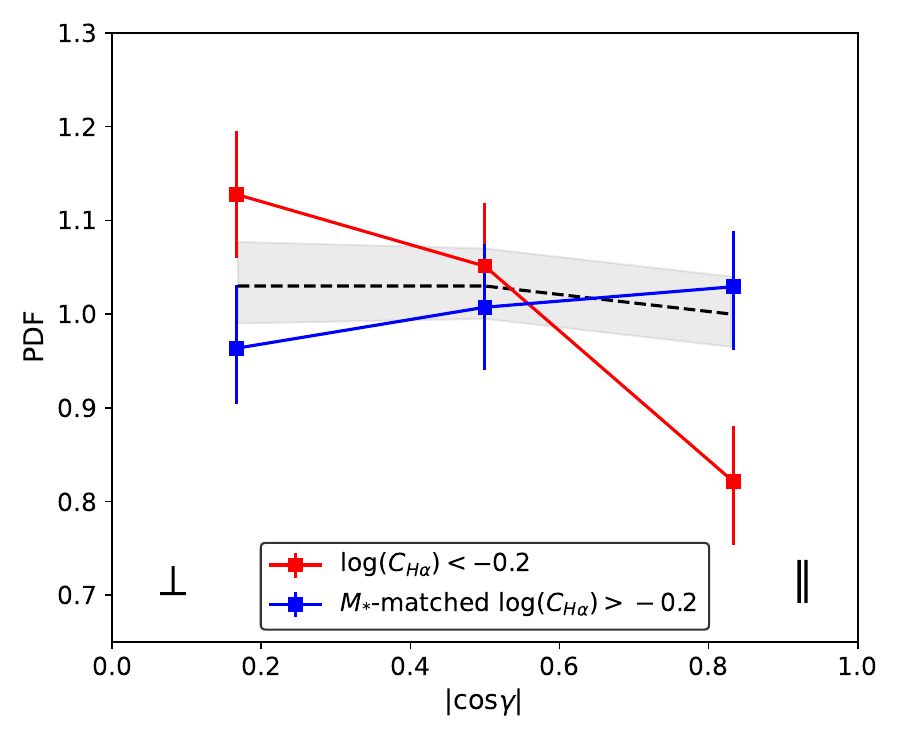}}
    \caption{PDF of galaxy spin–filament alignments for centrally-concentrated star-forming galaxies (C-index $<-0.2$) and stellar mass-matched regular star-forming galaxies (C-index $>-0.2$). The PDF is skewed towards more parallel spin–filament alignments for regular star-forming galaxies (blue) and more perpendicular spin–filament alignments for centrally-concentrated star-forming galaxies (red).}
    \label{Mass_matched_spin_flips}
\end{figure}

We further verify that the spin–filament alignment trends are not driven by the higher stellar masses of galaxies in dense environments by repeating the analysis matching the stellar-mass distributions of regular star-forming galaxies to those of centrally-concentrated systems. The tendency for centrally-concentrated star-forming galaxies to exhibit perpendicular spin–filament alignments remains present with respect to mass-matched regular star-forming galaxies showing a more parallel trend, as illustrated in Figure~\ref{Mass_matched_spin_flips}. The two |cos($\gamma$)| distributions also remain statistically significantly different ($p_{\rm 2KS}=0.038$), highlighting that the observed alignment trends are not driven by stellar mass selection effects.

\bsp	
\label{lastpage}
\end{document}